# Indoor microbiome, environmental characteristics and asthma among junior high school students in Johor Bahru, Malaysia

Xi Fu[1], Dan Norbäck[2], Qianqian Yuan[3,4], Yanling Li[3,4], Xunhua Zhu[3,4],Yiqun Deng[3,4], Jamal Hisham Hashim[5,6], Zailina Hashim[7], Yi-Wu Zheng[8], Xu-Xin Lai[8], Michael Dho Spangfort[8], Yu Sun[3,4] *

[1]Department of Occupational and Environmental Health, School of Public Health, Sun Yat-sen University, Guangzhou, PR China.

[2]Occupational and Environmental Medicine, Dept. of Medical Science, University Hospital, Uppsala University, 75237 Uppsala, Sweden.

[3]Guangdong Provincial Key Laboratory of Protein Function and Regulation in Agricultural Organisms, College of Life Sciences, South China Agricultural University, Guangzhou, Guangdong, 510642, PR China.

[4]Key Laboratory of Zoonosis of Ministry of Agriculture and Rural Affairs, South China Agricultural University, Guangzhou, Guangdong, 510642, PR China.

[5]United Nations University-International Institute for Global Health, Kuala Lumpur, Malaysia,

[6]Department of Community Health, National University of Malaysia, Kuala Lumpur, Malaysia

[7]Department of Environmental and Occupational Health, Faculty of Medicine and Health Sciences, Universiti Putra Malaysia, UPM, Serdang, Selangor, Malaysia

[8]Asia Pacific Research, ALK-Abello A/S, Guanzhou, China

*To whom corresponds should be addressed: Yu Sun sunyu@scau.edu.cn,

**Abstract**

Indoor microbial diversity and composition are suggested to affect the prevalence and severity of asthma, but no microbial association study has been conducted in tropical countries. In this study, we collected floor dust and environmental characteristics from 21 classrooms, and health data related to asthma symptoms from 309 students, in junior high schools in Johor Bahru, Malaysia. Bacterial and fungal composition was characterized by sequencing 16s rRNA gene and internal transcribed spacer (ITS) region, and the absolute microbial concentration was quantified by qPCR. In total, 326 bacterial and 255 fungal genera were characterized. Five bacterial (*Sphingobium*, *Rhodomicrobium*, *Shimwellia*, *Solirubrobacter*, *Pleurocapsa*) and two fungal (*Torulaspora* and Leptosphaeriaceae) taxa were protective for asthma severity. Two bacterial taxa, *Izhakiella* and *Robinsoniella*, were positively associated with asthma severity. Several protective bacterial taxa including *Rhodomicrobium*, *Shimwellia* and *Sphingobium* has been reported as protective microbes in previous studies, whereas other taxa were first time reported. Environmental characteristics, such as age of building, size of textile curtain per room volume, occurrence of cockroaches, concentration of house dust mite allergens transferred from homes by the occupants, were involved in shaping the overall microbial community but not asthma-associated taxa; whereas visible dampness and mold, which did not change the overall microbial community for floor dust, decreased the concentration of protective bacteria *Rhodomicrobium* ($\beta$=-2.86, p=0.021) of asthma, indicating complex interactions between microbes, environmental characteristics and asthma symptoms. Overall, this is the first indoor microbiome study to characterize the asthma-associated microbes and their environmental determinant in tropical area, promoting the understanding of microbial exposure and respiratory health in this region.

## Introduction

Asthma prevalence has been rising globally in the past few decades (Eder, Ege et al. 2006). It is suggested that the changing of life style and microbial exposure during the industrialization and urbanization process are associated with the increasing prevalence of asthma symptoms (Bello, Knight et al. 2018). Nowadays, more people live in city than rural area and they spend most of the time in the indoor environment (Klepeis, Nelson et al. 2001), thus it is necessary to identify the beneficial and risk exposure in various indoor environment. Progresses have been made in the past few years that several culture-independent microbiome studies revealing the association between indoor microbial exposure and human respiratory health. It was reported that high bacterial richness in homes of the farm area protected against childhood asthma compared with urban families (Ege, Mayer et al. 2011). Similarly, high diversity of fungal exposure is protective for childhood asthma development (Dannemiller, Mendell et al. 2014). However, there are also studies suggest that the asthma prevalence is related to the abundance of specific taxa rather than microbial richness (Kirjavainen, Karvonen et al. 2019). The number of microbiome surveys is limited and they adopted various study designs, sampling strategies and technique applications, thus the identified microbes generally varied between studies. For example, two microbiome studies used absolute quantification approaches identified only one protective or risk microbe for asthma symptoms (Dannemiller, Gent et al. 2016, Pekkanen, Valkonen et al. 2018), but one study in United States using relative abundance from 16s rRNA identified a few hundreds of potential associated microbes for inner-city children (O'Connor, Lynch et al. 2018). Furthermore, current microbiome studies are all from middle and high latitude areas, mainly from developed countries in Europe and United States. Indoor microbial composition is geographically patterned, especially for fungi, that significant community variation can be detected across different climate, latitude and geographic regions (Amend, Seifert et al. 2010, Barberan, Dunn et al. 2015). The associated-microbes identified in middle and high latitude provides little information regarding the microbial exposure and health in the tropical area. Thus, it is necessary to conduct indoor microbiome exposure and health effect for respiratory health in this region.

There are many epidemic studies on and childhood asthma, which usually have groups as 2-3 years, 3-5 years, and older than 5 years (Castro-Rodriguez, Forno et al. 2016, de Benedictis and Attanasi 2016, Bao, Chen et al. 2017). Most of the studies focus the groups younger than 5 years, and only a small amount of studies were conducted in adolescence. Some epidemic studies identified a list of common risk factors for asthma in adolescence, including furry per sensitization, early airway obstruction, parental rhinitis and asthma, the concentration of mold and endotoxins (Cai, Hashim et al. 2011, Norback, Markowicz et al. 2014, Norback, Hashim et al. 2017). In addition, perinatal familial stress, extreme preterm birth, and low birth weight may increase asthma risk during puberty (de Benedictis and Bush 2017). There are no studies investigating indoor microbiome in junior high schools and the association with asthma symptoms.

In this study, we conducted the first microbiome survey in a tropical area to screen protective and risk microbes associated with asthma symptoms. The aim of this study is to: 1) characterize microbiome composition in the floor dust from 21 classrooms in 7 randomly selected junior high schools in Johor Bahru, Malaysia; 2) quantitatively analyze the association between microbial taxa and prevalent of asthma; 3) identify indoor environmental characteristics influencing microbial community and asthma-associated taxa in the classrooms.

## Materials and Methods

Floor dust were collected from 8 junior high schools in Johor Bahru, Malaysia, 4 classes in each school. In total 32 dust samples were collected, but 11 of them failed to amplify adequate DNA for amplicon sequencing thus only dust samples of 21 classes could be sequenced. Health data were collected by self-reported questionnaires from 15 randomly selected students in each class. The ethical permission was approved by the Medical Research and Ethics Committee of the National University of Malaysia, and all participants gave their informed consent.

### Assessment of Health Data

Questions about doctor diagnosed asthma and current asthma was obtained from the European Community Respiratory Health Study (ECRHS). The questions included asthma symptoms and related information during last 12 months, including wheeze, breathlessness during wheeze, feeling of chest tightness, shortness of breath during rest, shortness of breath during exercise, woken by attack of shortness of breath, ever had asthma, attack of asthma, and current asthma medication use.

A validated asthma score including eight items were calculated to measure asthma severity (Pekkanen, Sunyer et al. 2005) were calculated, and then re-defined as 0, 1, 2, >=3. Questions about current smoking and parental asthma/allergy were also included. Details about the questions were described in previous study (Norback, Markowicz et al. 2014).

## Dust Sampling and Building Inspection

Floor dust in the classroom were collected by a 400 W vacuum cleaner with a dust sampler (ALK Abello, Copenhagen, Denmark) through a Milipore filter (pore size 6 μm). The total vacuum sampling procedure lasted 4 minutes, 2 minutes on the floor and 2 minutes on other surfaces above the floor like chairs and desks. Each classroom was divided into corridor part and window part, which were sampled separately into two samplers. The dust was then sieved in the lab, through a 0.3-mm mesh screen to fine dust, and were stored in the freezer at -80˚C. In this study, dust from the two parts of the classroom were combined. Environmental characteristics, including relative humidity, indoor $CO_2$ and outdoor $NO_2$ concentration, size of curtain, concentration of house dust mite and cockroach allergen, were measured, and information about the construction year, visible dampness and mold were noted (Norback, Markowicz et al. 2014).

## DNA extraction and sequencing

Total genome DNA were extracted for sequencing by Soil DNA Kit for all dust samples with bead beating and spin filter technology. A separate extraction of 10mg dust were conducted for real-time PCR. DNA quality and concentration was analyzed with a NanoDrop One spectrophotometer. Amplicons were generated by primers of v3 and v4 regions on the 16s ribosome RNA (16s rRNA) gene for bacteria, and internal transcript space 2 (ITS2) region for fungi.

Bioinformatics analysis and statistics

The forward and reverse reads were joined and assigned to samples by barcoding information, and the quality filter was set as sequence length >=200bp. The sequences were then assigned to operational taxonomic units (OTUs) with a sequence similarity of 97% and annotated against Silva and Unite database, respectively. Principle component analysis (PCoA) and Adonis analysis were performed to assess the influence of environmental characteristics to microbial richness and composition based on the distance matrix including Unifrac distance and Bray-Curtis analysis. Analyses were mainly conducted with the Quantitative Insights Into Microbial Ecology (QIIME, v1.8.0) platform (Caporaso, Kuczynski et al. 2010, Lawley and Tannock 2017). Two level hierarchical ordinal regression models were performed to analyze associations between microbial richness (number of observed taxonomic units, OTUs) and asthma score, and between quantity of single bacterial or fungal phylum, class, and genus (in log10 format) and asthma score. The latter analysis only included microbial taxa presented in at least five classrooms. In all analyses, gender, race, smoking, and parental asthma/allergy were included as adjustment. Parallel line assumption test were performed for the ordinal regression models, and those which violated the parallel line assumption were then calculated in a multi-nominal regression model. Association of single environmental characteristics with asthma score were assessed by hierarchical ordinal regression model. All hierarchical models and parallel line test were conducted by StataSE 15.0 (StataCorp LLC), and other statistics were conducted with IBM SPSS software 21.0 (IBM). Association between environmental characteristics and microbial richness and community variation were conducted by Adonis in R.

## Results

The response rate of the questionnaire was 96% (n=462). The students were aged from 14 to 16 years, and 52% were girls. In total, 309 (66.9%) of the participants were included, consisted of Malay (43%), Chinese (42%), and Indian (15%). The detailed demographic data were described in previous studies (Norback 2014, Cai 2011). The prevalence of asthma symptoms and asthma score is presented in Table 1.

*Sequencing statistics and microbial taxa*

The bacterial 16s rRNA dataset was rarefied to the depth of 27,000 reads for each samples, and fungal ITS was rarefied to 35,000 reads. The rarefaction curves indicate the sequencing depth is deep enough to capture the majority of operational taxonomic units (OTUs) in the floor dust (Figure S1). In total, 895 bacterial and 1,512 fungal OTUs were obtained, and distinct distribution patterns were observed. For bacteria, 36.8% of bacterial OTUs were presented in all samples, whereas only 10.3% of fungal OTUs were presented in all samples and approximately half of the OTUs were presented in ten or less samples (Figure 1A and 1B). The result suggests that many fungal taxa are presented in a few classrooms and are more locally distributed compared to bacterial taxa.

The major phylum included Proteobacteria (35.0 ± 7.3%, mean and standard deviation), Actinobacteria (21.2 ±6.3%), Cyanobacteria (17.6 ±7.3%) and Firmicutes (17.3 ±8.6%; Figure 1C and Table S1). The top genus mainly included environmental taxa such as *Bacillus* (4.2±5.8%), *Paracoccus* (3.2±1.2%), *Sphingomonas* (2.8±0.9%) and *Saccharopolyspora* (2.5±2.0%), and human skin taxa *Staphylococcus* (3.4±2.5%; Figure 1D, S2 and Table S2). The fungal phylum was dominated by Ascomycota (72.5±11.6%), followed by Basidiomycota (17.8±8.9%; Figure 1E). The top fungal genus included common mold taxa such as *Aspergillus* (16.6±7.9%), *Penicillium* (10.2±15.7%) and *Cladosporium* (7.8±8.2%), as well as outdoor environmental fungi such as *Hortaea* (8.0±7.5%), *Wallernia* (6.6±9.2%) and *Emericella* (4.1±9.6%) (Figure 1F, S3 and Table S3 and S4).

*Environmental characteristics associated with overall microbial richness/composition*

In this study, we collected eight environmental characteristics and tested their association with overall microbial richness and composition (Table S5 and S6). High concentration of house dust mite and cockroach allergens and higher textile factor (size of textile curtain per room volume) reduced the number of fungal observed OTUs (Adonis, $p < 0.05$; Table S5). The environmental characteristics had no associations with bacterial richness. Bacterial community composition was structured by the concentration of house dust mite allergens and textile factor in the classroom; fungal community was structured by age of building, concentration of house dust mite and cockroach allergens (Adonis, $p < 0.05$; Table 2). The results indicating the importance of textile material, occurrence of cockroaches in the indoor environment, and the concentration of house dust mite allergens transferred from homes in shaping the overall microbial composition. The role of textile factor in structuring microbial community was also illustrated by the principle coordinate analysis (PCoA) (Figure 1G and 1H). The bacterial composition was deviating along the first axis for classrooms with large or small curtain size per room volume.

*Identifying protective and risk microbes for asthma*

The prevalence of asthma symptoms is presented in Table 1. Asthma score was calculated based on all eight symptoms to represent the severity of asthma. There were no associations between the number of OTUs within the major phylum and class and asthma severity, suggesting microbial richness in the indoor environment was not significantly affecting asthma symptoms (Table S6). Although Proteobacteria (95% CI 0.92-1.01, p=0.09) and Cyanobacteria (0.95-1.0, p=0.07) showed marginally protective associations with the asthma score.

We screened associations between absolute bacterial and fungal quantity for genus and asthma score with a hierarchical ordinal regression model. To be conservative, p value <0.01 was set as cutoff to screen associated microbes. Six bacterial genera were negatively associated with asthma score (p<0.01), including *Sphingobium*, *Rhodomicrobium and Shimwellia* in Proteobacteria, *Solirubrobacter* in Actinobacteria, *Pleurocapsa* in Cyanobacteria, and JGI_0001001_H03 in Acidobacteria. Two genera, including *Izhakiella* in Proteobacteria and *Robinsoniella* in Firmicutes were positively associated with asthma score (Table 3). Two fungal genera in Ascomycota phylum were negatively associated with asthma score (p<0.01) (Table 3), including *Torulaspora,* and an unidentified genus in Leptosphaeriaceae family. The model for *Robinsoniella* and Leptosphaeriaceae violated the parallel line test (p<0.01), and the associations were then assessed by multi-nominal regression. Associations were observed between *Robinsoniella* and asthma score 0 to 1 (RRR=1.34, p<0.0001) and 0 to 2 (RRR=1.39, p=0.006), and between Leptosphaeriaceae and asthma score 0 to 1 (RRR=0.51, p=0.0001).

Recent studies claimed that absolute abundance approaches for microbial quantification should be used to link correct microbes to phenotypes and quantitative features, and the relative abundance approach can produce some erroneous identification and false positive results (Dannemiller, Mendell et al. 2014, Vandeputte, Kathagen et al. 2017). However, very few studies compare the absolute abundance with relative abundance in indoor microbiome survey studies. In addition to absolute abundance, we also conducted the association analysis between microbial relative abundance and asthma score with the same regression model, and found large variations between the two approaches. Among the eight associated microbes identified by absolute quantification, only three were identified by the relative abundance approach. Two additional genera, including *Wolbachia* and *Nocardiopsis*, were identified by the relative approach (Table S7). Similarly, only one fungal genus was identified by the relative abundance (Table S8).

*Environmental characteristics associated with protective microbes*

We investigated the associations between the environmental characteristics and the protective or risk microbes of asthma, and found that although the indoor dampness/visible mold was not a significant characteristic changing the overall indoor microbial composition for settled dust, it decreased the concentration of protective microbes, including *Rhodomicrobium* (β=-2.86,

p=0.021) and *Solirubrobacter* (β=-1.62, p=0.07). Thus, high indoor dampness can not only increase the prevalence of asthma by releasing submicron-sized cellular fragments and Microbial Volatile Organic Compounds (MVOCs) (Nevalainen, Taubel et al. 2015), it could also affect the respiratory health of occupants by suppressing the abundance of protective bacteria of asthma. To our knowledge, this is a new finding.

## Discussion

We identified 326 bacterial and 255 fungal genera from the indoor floor dust in seven junior high schools of Johor Bahru, Malaysia. Eight microbial taxa were quantitatively negatively associated with asthma severity, and two microbial taxa were positively associated. Visible indoor dampness and mold were not involved in shaping the overall microbial composition, but were negatively associated with the concentration of protective bacteria.

### Advantages and limitations

This is the first study to investigate the association between bacterial and fungal taxa and asthma symptoms in a tropical region. The study applied culture-free high-throughput sequencing and quantitative PCR to characterize the absolute concentration of microbial exposure in the classroom environment for adolescence in junior high schools. We systematically investigated the correlation between microbial exposure, environmental characteristics and asthma symptoms, revealing the complex relationship between these factors. There are also some limitations in our study. We used amplicon sequencing to characterize microbial composition in settle dust. Due to the technical limitation of amplicon sequencing, we can only identify the microbes down to the genus level, rather than more taxonomically resolved species and strain level. It is common that species within a genus or even strains from the same species could have different virulent factors thus posing different health effect for human. Thus, more taxonomically resolved technique, such as shotgun metagenomics, will improve the identification accuracy for future indoor microbiome survey. We identified several genus quantitatively associated with an asthma score, but due to the limitation of the cross-sectional study design, we can only report the association instead of a conclusion of causal effect. Further longitudinal studies are needed to disentangle the causal effect and temporal dynamic of the indoor microbiome.

In our study, we observed six bacterial genera protectively associated with asthma score ($p<0.01$) among students, including *Sphingobium*, *Rhodomicrobium*, *Shimwellia* in Proteobacteria, *Solirubrobacter* in Actinobacteria, *Pleurocapsa* in Cyanobacteria, and JGI_0001001_H03 in Acidobacteria. Except JGI_0001001_H03. The protective effect of these taxa has been previously reported in a microbiome study from farm and non-farm rural homes in Finland and Germany. In our study, the relative abundance of family Hyphomicrobiaceae (including *Rhodomicrobium*), Enterobacteriaceae (including *Shimwellia*), Sphigomonadaceae (including *Sphingobium*), the class Thermoleophilia (including *Solirubrobacter*), and the phylum Cyanobacteria (including *Pleurocapsa*) were higher in the farm home environment, which had protective effect for asthma symptoms (Kirjavainen 2019). The consistent result in tropical and European countries indicates a possible universal protective effect of these taxa across large geographic regions and in various climate conditions.

The other associated microbes identified in this study were not reported to be associated with respiratory health in previous studies. It is possible that the presence of these microbes is geographically restricted in tropical areas. *Izhakiella* is a newly identified genus, and recently isolated form mired bug and and Australian desert soil (Ji, Tang et al. 2017). The genus of *Robinsoniella* belongs the class of Clostridia. We found no research articles about this genus, but many other taxa in Clostridia class associated with asthma and human health. For example, *Clostridium cluster XI* in home environment was shown to be protectively associated with asthma prevalence among adults (Pekkanen, Valkonen 2018). Several families in Clostridia, including Phascolarctobacterium, Mogibacterium and Proteiniclasticum, were more abundant in rural farm homes, where there were lower asthma prevalence and healthier indoor microbiomes as compared to non-farm rural homes (Kirjavainen, 2019). In addition, *Clostridium butyricum* was suggested to be a potential therapeutic microbe combined with specific immunotherapy for asthma treatment (Clostridium butyricum in combination with specific immunotherapy converts antigen-specific B cells to regulatory B cells in asthmatic patients). However, harmful effect of Clostridia taxa has also been reported, such as *Clostridium difficile*, which can cause serious diarrhea to life-threatening colitis (Borali and De Giacomo 2016, Smits, Lyras et al. 2016).

Previous culture-dependent studies have reported *Aspergillus*, *Penicillium*, *Alternaria* and Cladosporium as risk fungal taxa for asthma symptoms (Sharpe, Bearman et al. 2015). However, very few studies applies culture-independent approaches to systematically screen fungal microbes studies for asthma symptoms. Dannemiller et al. identified *Volutella* was positively and *Kondoa* was protectively associated with asthma severity among atopic and nonatopic children in the Northeast of United States (Dannemiller, Gent et al. 2016). In this study, we identified one protective fungi taxa, *Torulaspora*. *Torulaspora delbrueckii* has been shown to have probiotic potential that can be used as supplement in food production to regulate intestinal response and promote human health (Zivkovic, Cadez et al. 2015). The family of Leptosphaeriaceae has not been previously reported to be associated with human health.

Among the building characteristics related to asthma, only indoor dampness/visible mold were associated with genus related to asthma. Indoor dampness/visible mold was negatively associated with bacterial genera protective to asthma severity, *Solirubrobacter* and *Rhodomicrobium*. This is new to our knowledge. Dampness and mold have been proved as risk factors for respiratory health, including asthma, for a few years (Quansah, Jaakkola et al. 2012, Castro-Rodriguez, Forno et al. 2016). A recent study has reported that indoor dampness and mold increase onset of asthma symptoms and reduce remission from asthma among adults (Wang, Pindus et al. 2019). Previous studies on dampness and mold in buildings established the direct association between mold species, fungal cellular fragment and MVOCs and related airway inflammation (Nevalainen, Taubel et al. 2015, An and Yamamoto 2016, Zhang, Reponen et al. 2016). Our results suggest that, despite the direct harmful effect from fungi, mold growth may suppress the concentration of beneficial bacteria that are protective for asthma symptoms. It has been reported that in floor dust, the absolute concentration of most mold taxa, including *Aspergillus*, *Penicillium* and *Alternaria*, keeps increasing with elevated humidity, whereas the concentration of certain bacterial taxa, including Pasteurellaceae, *Prevotella* and *Cytophaga*, decreases with elevated humidity (Dannemiller, Weschler et al. 2017). However, the dynamics of fungal and bacterial growth is a complex issue, the detailed interaction among microbes are still unclear. As the majority of fungal and bacterial species are non-culturable, new study designs such as shotgun metagenomics sequencing strategy combined with in silico growth rate analysis, such as tools like GRiD (Emiola and Oh 2018), holds promising solution for the issue.

In our study, we used absolute taxonomic quantities to assess the associations with asthma score, while some previous studies used relative abundance (O'Connor, Lynch et al. 2018, Kirjavainen, Karvonen et al. 2019). An issue of relative abundance is that the abundance change of one of the taxa will lead to abundance changes of all other taxa, which may lead to over-identification or mis-identification. O’Connor et al. have identified 201 risk and 171 protective microbes from household dust associated with prevalent asthma among children (O'Connor, Lynch et al. 2018). Among these microbes, some common human skin bacteria have been identified as risky microbes for childhood asthma, including *Staphylococcus, Corynebacterium, Haemophilus* and *Sphingomonas*, but they are unlikely to be risk agents since these taxa are universally present around human occupants. One study on quantification profiling of gut microbiome has reported that absolute abundance of microbes significantly differ from the rank by relative abundance, which affects the result of association analysis for phenotypes (Vandeputte, Kathagen et al. 2017). In this study, only 3 bacteria identified by relative approach were consistent with absolute approach, indicating the discrepancy between the two approaches. From our results, we observed that most the identified taxa related to health were minor taxa with a relative abundance <0.2% in all samples. The relative abundance of these minor microbes can be impacted by the variation of dominant microbes, and their concentration is more appropriately presented by absolute approach.

**Conclusion**

In this study, we reported a list of minor bacterial and fungal taxa that were associated with asthma severity in classrooms in junior high school students in Johor Bahru, Malaysia. Environmental characteristics associated with overall microbial community were not associated with the protective or risk microbes, but indoor dampness/ visible mold decreased the quantity of bacteria protective to asthma. This is the first study to reveal the complex interaction between microbiome, environmental characteristics and asthma symptoms in a tropical area. The study contributes to new knowledge on how to promote the establishment of a healthy building microbiome in this region.

Table 1. Prevalence of asthma symptoms and asthma score

| Symptoms | Number (n=309) | Prevalence (%) |
|---|---|---|
| Wheeze and breathlessness during wheeze | 20 | 6.56 |
| Feeling of chest tight | 16 | 5.18 |
| Attack of shortness of breath during rest | 28 | 9.1 |
| Attack of shortness of breath during exercise | 114 | 36.9 |
| Woken by attack of shortness of breath | 23 | 7.4 |
| Ever asthma | 39 | 12.6 |
| Attack of asthma | 7 | 2.3 |
| Current medication for asthma | 11 | 3.6 |
| Asthma score | | |
| 0 | 148 | 47.9 |
| 1 | 101 | 32.7 |
| 2 | 28 | 9.1 |
| >=3 | 32 | 10.3 |

Table 2. Association between outdoor/indoor characteristics and microbial community variation based on weighted unifrac and Bray-Curtis distance (beta diversity). P value was calculated based on 10,000 permutation bivariate Adonis analysis.

| | **Bacteria** | | **Fungi** | |
|---|---|---|---|---|
| | $R^2$ | p | $R^2$ | p |
| Relative humidity | 0.05 | 0.39 | 0.06 | 0.25 |
| Indoor $CO_2$ | 0.02 | 0.91 | 0.02 | 0.99 |
| Outdoor $NO_2$ | 0.09 | 0.06 | 0.04 | 0.6 |
| Building age | 0.06 | 0.29 | **0.1** | **0.03** |
| Visible indoor dampness/mold | 0.03 | 0.84 | 0.03 | 0.81 |
| Textile curtain factor | **0.16** | **0.005** | 0.09 | 0.06 |
| Concentration of house dust mite allergen in dust | **0.2** | **0.004** | **0.15** | **0.009** |
| Concentration of cockroach allergen in dust | 0.05 | 0.4 | **0.17** | **0.04** |

Table 3. Microbial taxa associated with asthma severity status in junior high school by ordinal regression analysis (p<0.01). Taxonomic information of the associated microbes were listed.

| Kingdom | Phylum | Class | Genus | GM, GSD (copies / g dust) | Range of relative abundance (%) | Number of classrooms with presence (N=21) | OR (95% CI) | p value |
|---|---|---|---|---|---|---|---|---|
| Bacteria | Proteobacteria | Alphaproteobacteria | *Sphingobium* | 156, 1.4 | 0.02-0.15 | 21 | 0.368 (0.187-0.724) | 0.004 |
| | | | *Rhodomicrobium* | 53, 7.5 | 0-0.13 | 17 | 0.837 (0.749-0.939) | 0.002 |
| | | Gammaproteobacteria | *Shimwellia* | 167, 2.0 | 0.01-1.65 | 21 | 0.560 (0.385-0.813) | 0.002 |
| | | | *Izhakiella* | 3, 8.0 | 0-0.13 | 5 | 1.187 (1.051-1.342) | 0.006 |
| | Actinobacteria | Thermoleophilia | *Solirubrobacter* | 66, 4.1 | 0-0.04 | 19 | 0.787 (0.672-0.923) | 0.003 |
| | Cyanobacteria | Oxyphotobacteria | *Pleurocapsa* | 140, 1.6 | 0.01-0.26 | 21 | 0.465 (0.275-0.788) | 0.005 |
| | Firmicutes | Clostridia | *Robinsoniella* | 17, 9.4 | 0-0.11 | 13 | 1.182 (1.054-1.326) | 0.004 |
| Fungi | Ascomycota | Saccharomycetes | *Torulaspora* | 2, 2.2 | 0-0.01 | 7 | 0.643 (0.476-0.870) | 0.004 |
| | | Dothideomycetes | f_Leptosphaeriaceae g_unidentified | 2, 2.5 | 0-0.12 | 7 | 0.672 (0.505-0.893) | 0.006 |

Significance level: p<0.01.

Sex, race, smoking, and parental asthma or allergy were adjusted for in the association analyses.

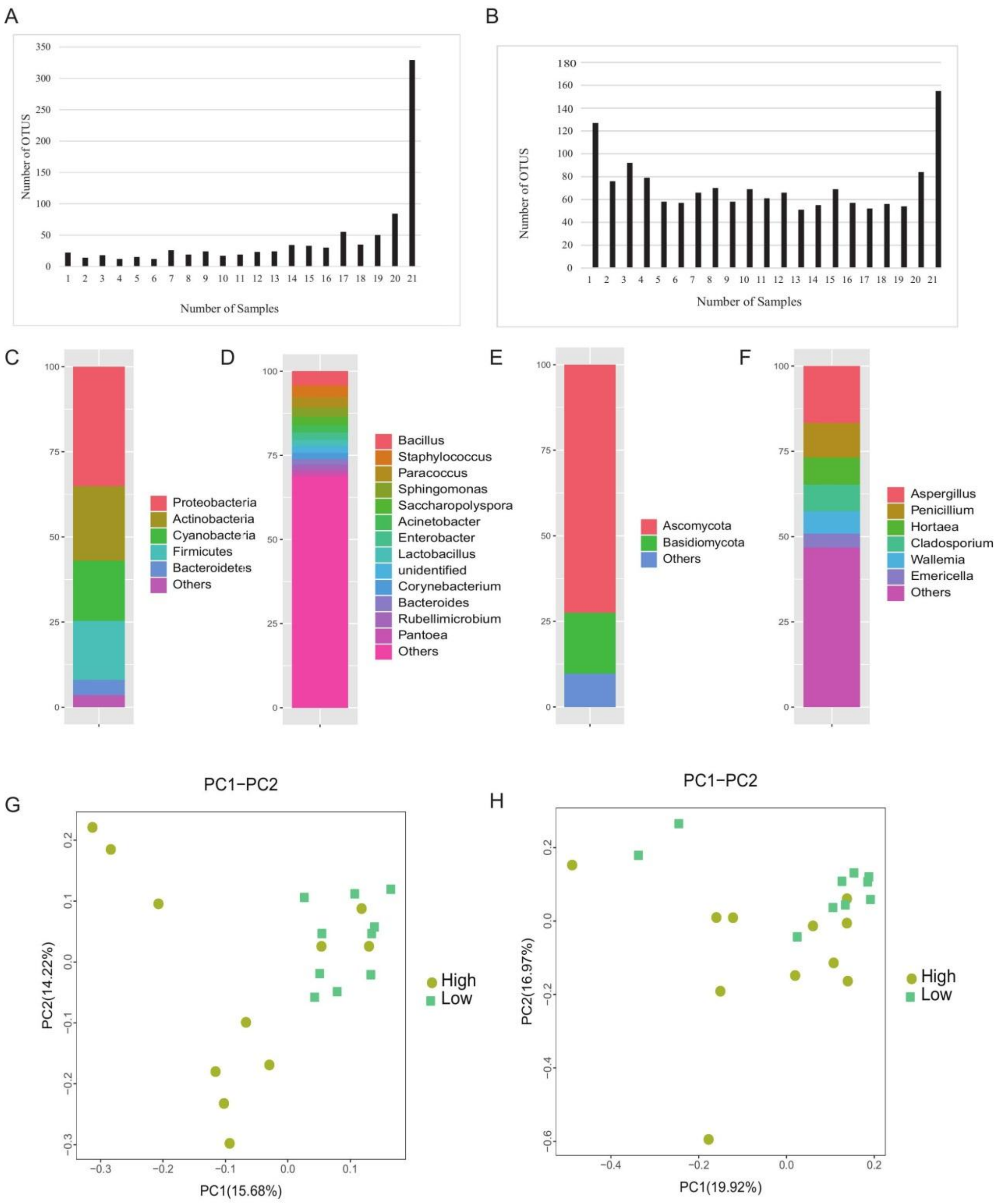

Figure 1. (A) Bacterial OTUs distribution. (B) Fungal OTUs distribution. (C) The major phylum of bacteria. (D) The top genus of bacteria. (E) The major phylum of fungi. (F) The top genus of fungi. (G) The principle coordinate analysis (PCoA) of bacteria structured by textile factor. (H) The principle coordinate analysis (PCoA) of fungi structured by textile factor.

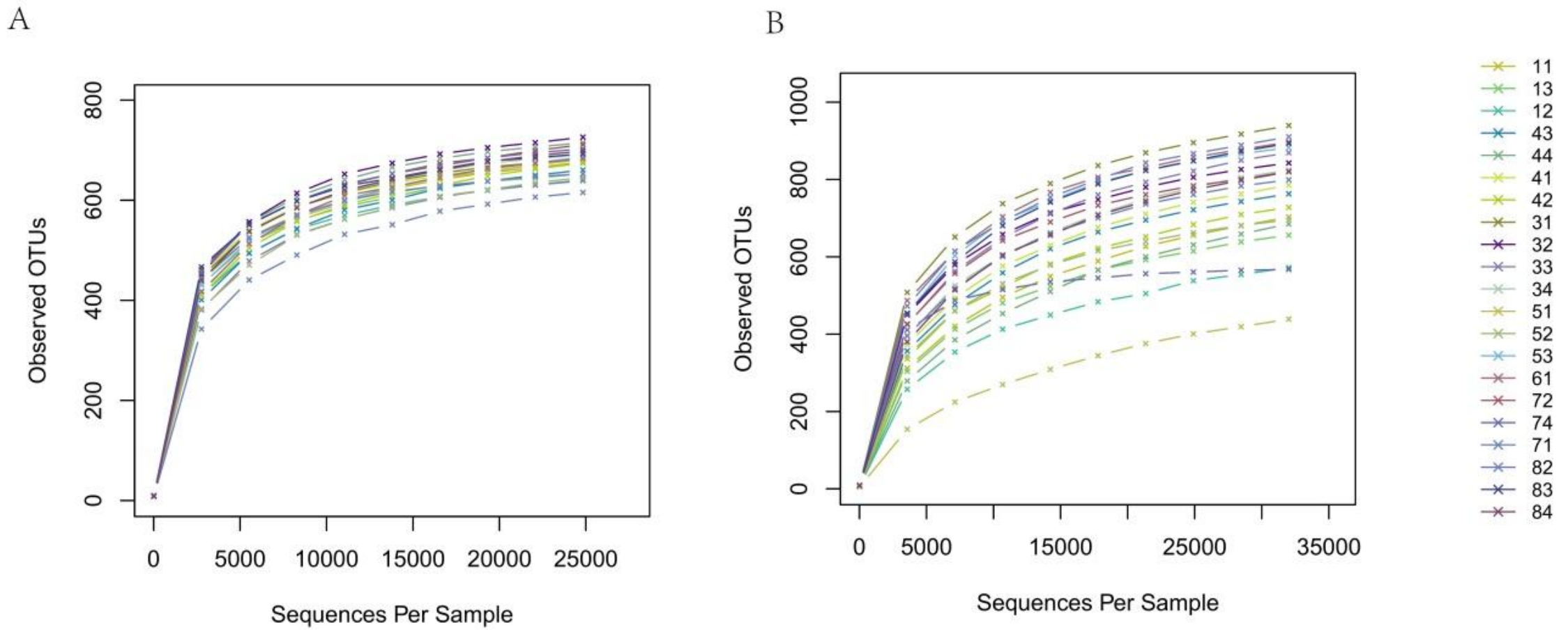

Figure S1. The rarefaction curves.

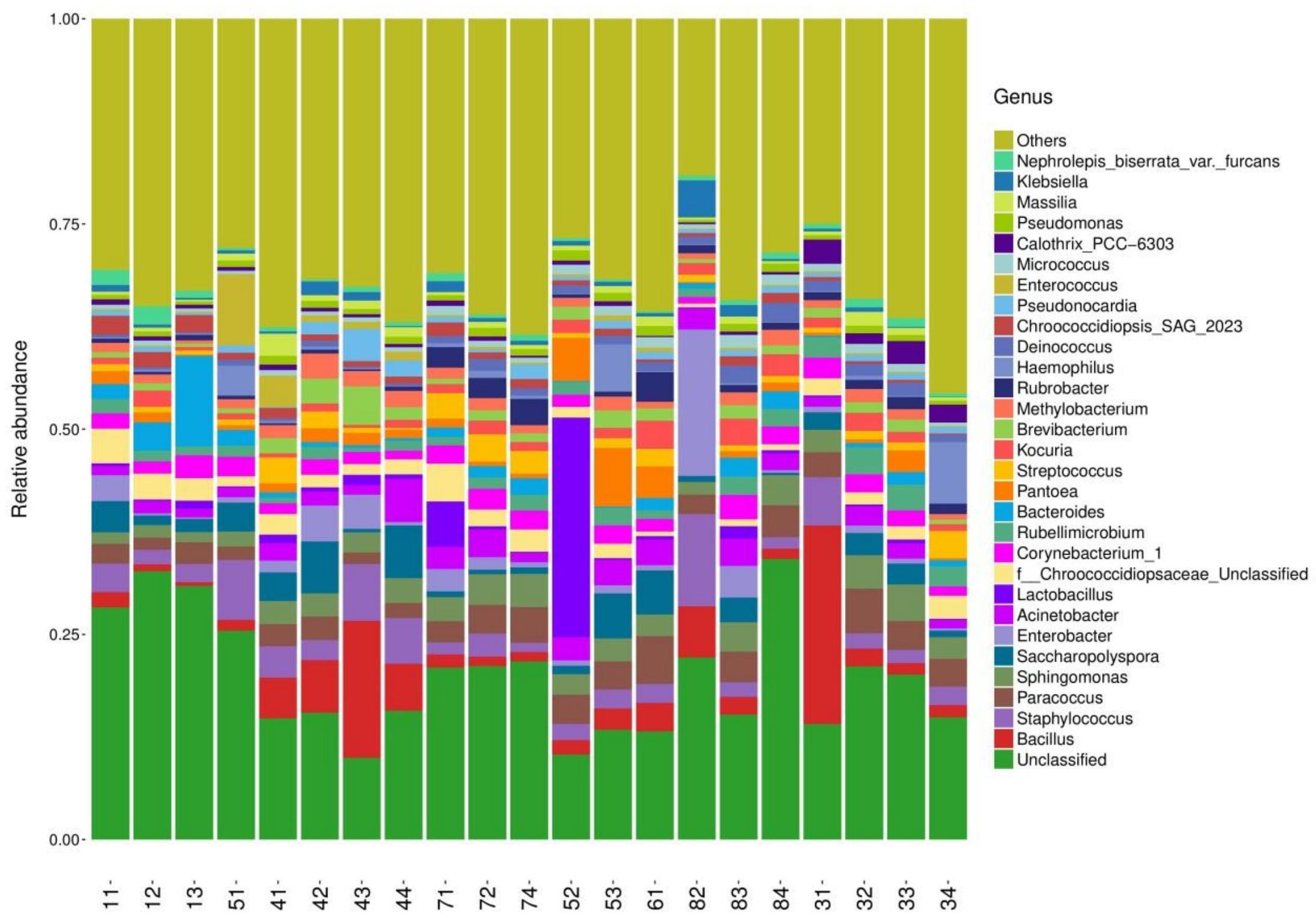

Figure S2. Bar of genus of bacteria in 21 samples.

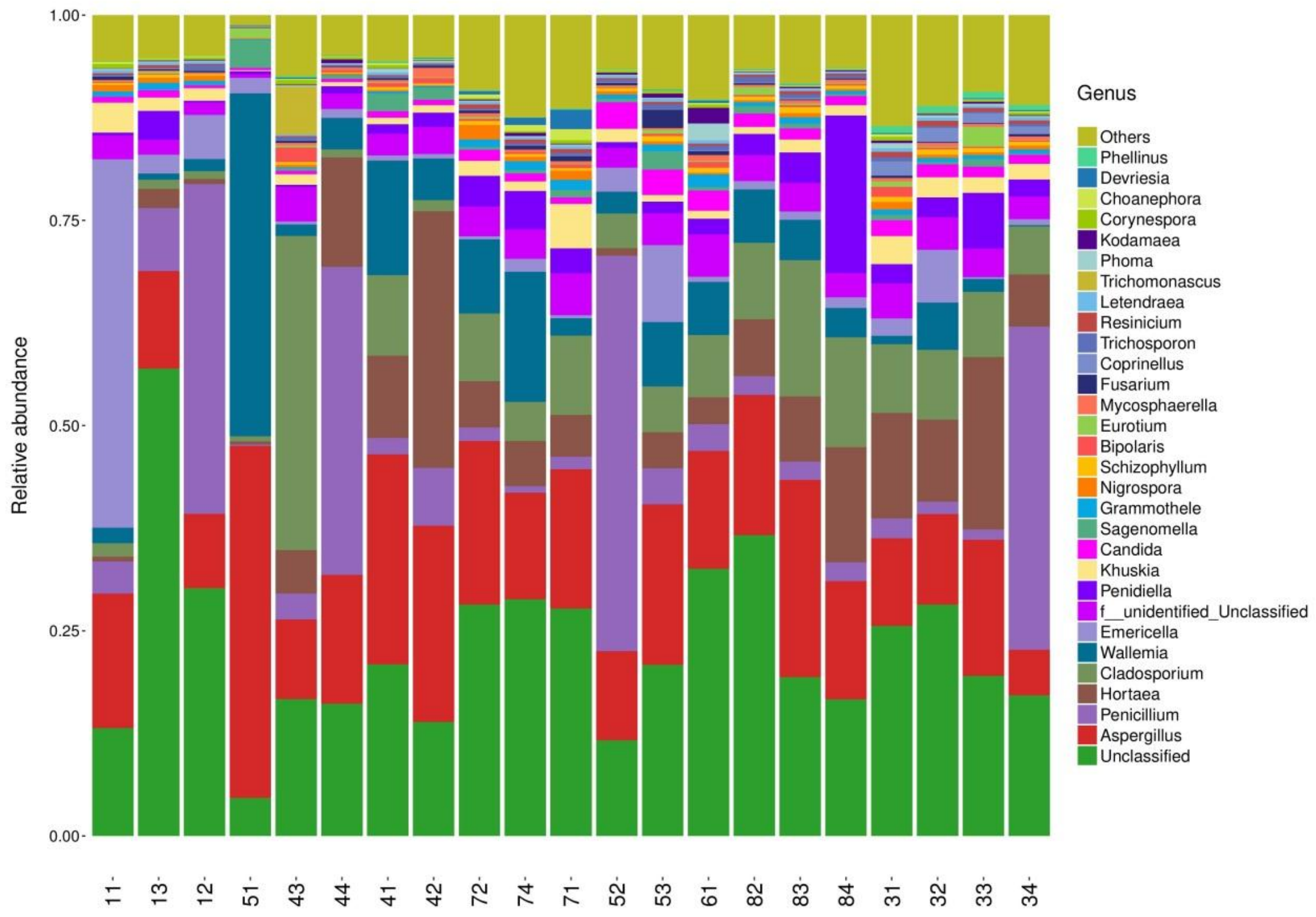

Figure S3. Bar of genus of fungi in 21 samples.

TableS1 Bacteria phylum abundance.

| Taxon | average abundance | standard deviation |
|---|---|---|
| Proteobacteria | 35.03524 | 7.285408 |
| Actinobacteria | 21.95524 | 6.337985 |
| Cyanobacteria | 17.62381 | 7.344191 |
| Firmicutes | 17.32333 | 8.616899 |
| Bacteroidetes | 4.511429 | 3.88505 |
| Deinococcus-Thermus | 1.494762 | 0.614318 |
| Acidobacteria | 0.760952 | 0.335915 |
| Chloroflexi | 0.589048 | 0.292522 |
| Fusobacteria | 0.162381 | 0.350598 |
| Gemmatimonadetes | 0.123333 | 0.106176 |
| Armatimonadetes | 0.11619 | 0.064224 |
| Unclassified | 0.107143 | 0.097373 |
| Patescibacteria | 0.104286 | 0.346635 |
| Verrucomicrobia | 0.052381 | 0.075359 |
| Epsilonbacteraeota | 0.03381 | 0.130172 |
| WPS-2 | 0.00381 | 0.004976 |
| Spirochaetes | 0.002381 | 0.007003 |
|  | 3.550951 |  |

TableS2 Bac genus abundance.

| Taxon | average abundance | standard deviation |
|---|---|---|
| Unclassified | 20.06381 | 7.264851 |
| Bacillus | 4.220476 | 5.811346 |
| Staphylococcus | 3.371429 | 2.545251 |
| Paracoccus | 3.131905 | 1.206837 |
| Sphingomonas | 2.760476 | 0.911359 |
| Saccharopolyspora | 2.5 | 2.010614 |
| Acinetobacter | 2.21 | 1.118356 |
| Enterobacter | 2.209048 | 3.850878 |
| Lactobacillus | 1.956667 | 5.736908 |
| f__Chroococcidiopsaceae_Unclassifie | 1.909524 | 1.041088 |
| Corynebacterium_1 | 1.77381 | 0.550141 |
| Bacteroides | 1.670952 | 2.302336 |
| Rubellimicrobium | 1.654286 | 0.753516 |
| Pantoea | 1.56381 | 1.778096 |
| Streptococcus | 1.45 | 1.072632 |
| Kocuria | 1.43 | 0.901604 |
| Brevibacterium | 1.419048 | 0.959041 |
| Methylobacterium | 1.32619 | 0.571222 |
| Rubrobacter | 1.116667 | 0.969924 |
| Haemophilus | 1.013333 | 2.024256 |
| Pseudonocardia | 0.963333 | 0.824902 |
| Deinococcus | 0.949524 | 0.585025 |
| Chroococcidiopsis_SAG_2023 | 0.938571 | 0.586313 |
| Enterococcus | 0.921429 | 1.990752 |
| Micrococcus | 0.819048 | 0.363619 |
| Calothrix_PCC-6303 | 0.808571 | 0.810138 |
| Pseudomonas | 0.801429 | 0.266407 |
| Massilia | 0.76381 | 0.57256 |
| Klebsiella | 0.747143 | 0.932063 |
| Nephrolepis_biserrata_var._furcans | 0.720476 | 0.498944 |
| Brachybacterium | 0.709524 | 0.362967 |
| Skermanella | 0.706667 | 0.392012 |
| Parabacteroides | 0.661905 | 0.868617 |
| Janibacter | 0.644286 | 0.369697 |
| Roseomonas | 0.640476 | 0.337868 |
| Actinomycetospora | 0.589524 | 0.256622 |
| Escherichia-Shigella | 0.575714 | 0.409836 |
| Truepera | 0.544286 | 0.183618 |
| MN_122.2a | 0.54381 | 0.28521 |
| Geodermatophilus | 0.53381 | 0.297497 |
| Curtobacterium | 0.508095 | 0.312404 |
| Scytonema_UCFS19 | 0.491429 | 0.683339 |
| Arthrobacter | 0.484762 | 0.735484 |
| Chroococcidiopsis_PCC_7203 | 0.461429 | 0.277043 |
| Nocardioides | 0.457143 | 0.241974 |
| Leptolyngbya_PCC-6306 | 0.455238 | 0.246427 |
| Nesterenkonia | 0.435714 | 0.371451 |
| CENA359 | 0.431905 | 0.28326 |
| Marmoricola | 0.428571 | 0.354363 |
| Clostridium_sensu_stricto_1 | 0.420952 | 0.264611 |
| Brevundimonas | 0.407619 | 0.162447 |
| Blastococcus | 0.406667 | 0.251044 |
| Enhydrobacter | 0.392857 | 0.216128 |
| Exiguobacterium | 0.374286 | 0.331927 |
| 1174-901-12 | 0.369048 | 0.152935 |
| Scytonema_VB-61278 | 0.367143 | 0.270576 |
| Gordonia | 0.359048 | 0.243493 |
| Hymenobacter | 0.356667 | 0.193864 |
| Dapisostemonum_CCIBt_3536 | 0.34619 | 0.338681 |
| Lachnoclostridium | 0.324286 | 0.617985 |
| Pseudokineococcus | 0.322381 | 0.325775 |
| Allorhizobium-Neorhizobium-Pararhiz | 0.32 | 0.215801 |
| Neisseria | 0.291905 | 0.665016 |
| f__uncultured_Unclassified | 0.283333 | 0.216295 |
| Ignatzschineria | 0.254286 | 0.810084 |
| Weissella | 0.253333 | 0.262723 |
| Kurthia | 0.249524 | 0.842659 |
| Stenotrophomonas | 0.243333 | 0.544282 |
| Mycobacterium | 0.244286 | 0.079093 |
| Proteus | 0.241429 | 0.669659 |
| Romboutsia | 0.241905 | 0.222298 |
| Glutamicibacter | 0.231429 | 0.346342 |
| Bryum_argenteum_var._argenteum | 0.225238 | 0.25901 |
| Scytonema_UTEX_2349 | 0.221905 | 0.355902 |
| f__uncultured_cyanobacterium_Unclas | 0.217619 | 0.277685 |
| Rothia | 0.205238 | 0.159738 |
| Mastigocladopsis_PCC-10914 | 0.20381 | 0.179902 |
| Shimwellia | 0.203333 | 0.441784 |
| Quadrisphaera | 0.202381 | 0.107931 |
| Aeromonas | 0.196667 | 0.170421 |
| Novosphingobium | 0.193333 | 0.092376 |
| f__Blastocatellaceae_Unclassified | 0.193333 | 0.11547 |
| Bryobacter | 0.191905 | 0.150686 |
| Desulfovibrio | 0.189524 | 0.384233 |
| Qipengyuania | 0.187143 | 0.122969 |
| Odoribacter | 0.181429 | 0.281377 |
| Spirosoma | 0.179048 | 0.130073 |
| Scytonema_PCC-7110 | 0.175714 | 0.135262 |
| Serratia | 0.169048 | 0.230432 |
| Barrientosiimonas | 0.170476 | 0.1059 |
| Aliterella_CENA595 | 0.169048 | 0.105826 |
| Gardnerella | 0.167143 | 0.507909 |
| Chryseobacterium | 0.167619 | 0.164375 |
| Catellicoccus | 0.164286 | 0.28067 |
| Lactococcus | 0.163333 | 0.189719 |
| Cellulosimicrobium | 0.159524 | 0.105758 |
| endosymbionts8 | 0.152857 | 0.149704 |
| Kosakonia | 0.150476 | 0.278253 |
| Actinomyces | 0.147143 | 0.231951 |
| Micromonospora | 0.147619 | 0.124535 |
| Alishewanella | 0.145714 | 0.097753 |
| Jeotgalicoccus | 0.145714 | 0.258777 |
| Moraxella | 0.144286 | 0.322127 |
| f__Corynebacteriaceae_Unclassified | 0.145238 | 0.483111 |
| Megamonas | 0.144286 | 0.386815 |
| f__Ilumatobacteraceae_Unclassified | 0.143333 | 0.088788 |
| f__Beijerinckiaceae_Unclassified | 0.129048 | 0.054213 |
| f__Unknown_Family_Unclassified | 0.128095 | 0.069973 |
| Haliangium | 0.12619 | 0.379005 |
| Capnocytophaga | 0.124286 | 0.490669 |
| Gemmatirosa | 0.122381 | 0.104876 |
| Hungatella | 0.121905 | 0.128749 |
| f__JG30-KF-CM45_Unclassified | 0.120476 | 0.069963 |
| Kytococcus | 0.120476 | 0.086283 |
| PMMR1 | 0.119048 | 0.059742 |
| Empedobacter | 0.118095 | 0.080039 |
| Salinicoccus | 0.115714 | 0.13006 |
| Modestobacter | 0.114286 | 0.109662 |
| Terriglobus | 0.115238 | 0.064778 |
| Marinococcus | 0.114762 | 0.127774 |
| Dolosigranulum | 0.115714 | 0.096466 |
| Prevotella_2 | 0.110952 | 0.456299 |
| f__uncultured_bacterium_Unclassifie | 0.104286 | 0.066601 |
| Amaricoccus | 0.102857 | 0.053865 |
| Leptolyngbya_ANT.L52.2 | 0.105238 | 0.097448 |
| Faecalibacterium | 0.102857 | 0.216221 |
| f__Saccharimonadaceae_Unclassified | 0.103333 | 0.346689 |
| f__Roseiflexaceae_Unclassified | 0.099524 | 0.100473 |
| Wolbachia | 0.097619 | 0.109403 |
| Aggregatibacter | 0.094762 | 0.381911 |
| Blautia | 0.094762 | 0.113297 |
| f__Caulobacteraceae_Unclassified | 0.095714 | 0.042611 |
| Chalicogloea_CCALA_975 | 0.093333 | 0.142945 |
| Fusobacterium | 0.092381 | 0.196237 |
| Providencia | 0.09 | 0.207171 |
| Microvirga | 0.091429 | 0.064752 |
| Veillonella | 0.089048 | 0.106673 |
| Ornithinimicrobium | 0.09 | 0.056921 |
| RB41 | 0.089048 | 0.071478 |
| Cellulomonas | 0.087143 | 0.073698 |
| Prauserella | 0.087619 | 0.214637 |
| f__Lachnospiraceae_Unclassified | 0.086667 | 0.204361 |
| Paeniclostridium | 0.083333 | 0.151537 |
| Peptostreptococcus | 0.08381 | 0.249609 |
| Alkanindiges | 0.081905 | 0.064391 |
| Dietzia | 0.082857 | 0.043491 |
| Prevotella_7 | 0.081905 | 0.144555 |
| Pseudocitrobacter | 0.080952 | 0.099242 |
| Lautropia | 0.081429 | 0.260563 |
| Piscicoccus | 0.077619 | 0.078607 |
| Bounagaea | 0.07619 | 0.148205 |
| Kineosporia | 0.074762 | 0.053723 |
| Halomonas | 0.074762 | 0.072223 |
| Cnuella | 0.07381 | 0.065075 |
| Peptoclostridium | 0.071905 | 0.160144 |
| Tissierella | 0.07 | 0.318496 |
| Porphyromonas | 0.068571 | 0.158565 |
| Sphingobium | 0.069048 | 0.079618 |
| Leptotrichia | 0.068095 | 0.16588 |
| Aureimonas | 0.069524 | 0.034275 |
| Granulicatella | 0.06619 | 0.057138 |
| Comamonas | 0.065238 | 0.047919 |
| Dermacoccus | 0.064286 | 0.036135 |
| Sandaracinobacter | 0.064286 | 0.043078 |
| Aliicoccus | 0.063333 | 0.037327 |
| Kineococcus | 0.063333 | 0.043397 |
| Eggerthella | 0.061429 | 0.062951 |
| Aetokthonos_AEL04 | 0.061905 | 0.070825 |
| Candidatus_Alysiosphaera | 0.06 | 0.036878 |
| Streptomyces | 0.059524 | 0.039176 |
| Lysinibacillus | 0.056667 | 0.084222 |
| f__Coleofasciculaceae_Unclassified | 0.057143 | 0.079633 |
| Adhaeribacter | 0.055714 | 0.218942 |
| Bryocella | 0.054762 | 0.026574 |
| f__Archangiaceae_Unclassified | 0.052381 | 0.062921 |
| Pleurocapsa_PCC-7327 | 0.053333 | 0.053041 |
| Rheinheimera | 0.053333 | 0.033066 |
| Akkermansia | 0.052381 | 0.075359 |
| Haloechinothrix | 0.051905 | 0.171977 |
| Abiotrophia | 0.05 | 0.161245 |
| [Ruminococcus]_torques_group | 0.049048 | 0.067743 |
| Saccharibacillus | 0.05 | 0.094128 |
| Flavisolibacter | 0.049048 | 0.041941 |
| Fictibacillus | 0.049048 | 0.024475 |
| Craurococcus | 0.048095 | 0.052021 |
| Altererythrobacter | 0.048571 | 0.035112 |
| Marinilactibacillus | 0.04619 | 0.164118 |
| Aerococcus | 0.04619 | 0.048423 |
| Sphingobacterium | 0.047143 | 0.045841 |
| Belnapia | 0.047619 | 0.016705 |
| Rikenella | 0.045238 | 0.074607 |
| Chakia_8 | 0.046667 | 0.094675 |
| Bifidobacterium | 0.045238 | 0.066379 |
| Shewanella | 0.04619 | 0.038533 |
| Eubacterium | 0.04381 | 0.058007 |
| Listeria | 0.042381 | 0.124776 |
| Actinoplanes | 0.042381 | 0.038588 |
| [Ruminococcus]_gnavus_group | 0.041429 | 0.091939 |
| Vittaria_lineata__[shoestring_fern] | 0.041905 | 0.035443 |
| Gemella | 0.041905 | 0.041787 |
| Rhodomicrobium | 0.04 | 0.043818 |
| Dialister | 0.040476 | 0.180928 |
| Achromobacter | 0.042381 | 0.027185 |
| f__Xanthobacteraceae_Unclassified | 0.040476 | 0.043529 |
| Defluviicoccus | 0.039048 | 0.027911 |
| f__A4b_Unclassified | 0.036667 | 0.055438 |
| Savagea | 0.036667 | 0.112798 |
| Lawsonella | 0.03619 | 0.028368 |
| Leptolyngbya_VRUC_135 | 0.035714 | 0.039316 |
| Devosia | 0.03619 | 0.026547 |
| Roseburia | 0.03619 | 0.028892 |
| Campylobacter | 0.03381 | 0.130172 |
| Microcoleus_SAG_1449-1a | 0.034286 | 0.059209 |
| Peptoniphilus | 0.033333 | 0.06491 |
| Wohlfahrtiimonas | 0.033333 | 0.106317 |
| Finegoldia | 0.034286 | 0.030095 |
| Nocardiopsis | 0.033333 | 0.053417 |
| Paenibacillus | 0.033333 | 0.054985 |
| f__Nostocaceae_Unclassified | 0.031905 | 0.039322 |
| Cedecea | 0.032381 | 0.03375 |
| Sphingoaurantiacus | 0.031905 | 0.028039 |
| Vibrio | 0.029524 | 0.051911 |
| Turicella | 0.03 | 0.084083 |
| [Eubacterium]_fissicatena_group | 0.029048 | 0.04979 |
| f__Ambiguous_taxa_Unclassified | 0.029524 | 0.021325 |
| Thauera | 0.028095 | 0.05335 |
| Megasphaera | 0.027143 | 0.057284 |
| Nakamurella | 0.027143 | 0.013836 |
| Pseudoxanthomonas | 0.026667 | 0.01278 |
| Luteimonas | 0.02619 | 0.020119 |
| Antricoccus | 0.02619 | 0.018835 |
| Microcoleus_Es-Yyy1400 | 0.024286 | 0.05259 |
| f__Caldilineaceae_Unclassified | 0.024762 | 0.033559 |
| YB-42 | 0.02381 | 0.028719 |
| Segniliparus | 0.02381 | 0.10684 |
| Bilophila | 0.024286 | 0.03641 |
| Anaerostipes | 0.02381 | 0.04544 |
| Alistipes | 0.02381 | 0.051524 |
| Cloacibacterium | 0.024286 | 0.021112 |
| Plesiomonas | 0.022381 | 0.086828 |
| Acidothermus | 0.021905 | 0.038421 |
| Cyanothece_PCC-7424 | 0.021905 | 0.02542 |
| Tatumella | 0.021905 | 0.02205 |
| Porphyrobacter | 0.020952 | 0.019211 |
| f__Euzebyaceae_Unclassified | 0.02 | 0.041231 |
| Vagococcus | 0.019048 | 0.038458 |
| Microseira_Carmichael-Alabama | 0.019048 | 0.022114 |
| Solirubrobacter | 0.019048 | 0.013002 |
| Granulicella | 0.018571 | 0.014243 |
| f__Rhizobiaceae_Unclassified | 0.017619 | 0.041341 |
| EcFYyy-200 | 0.019048 | 0.02427 |
| Solibacillus | 0.016667 | 0.031198 |
| Mucilaginibacter | 0.016667 | 0.037594 |
| Blastocatella | 0.016667 | 0.038123 |
| Ochrobactrum | 0.016667 | 0.020083 |
| Mesorhizobium | 0.015238 | 0.016619 |
| [Eubacterium]_coprostanoligenes_grc | 0.015714 | 0.01886 |
| Robinsoniella | 0.014762 | 0.024211 |
| GCA-900066225 | 0.014762 | 0.02272 |
| Terrisporobacter | 0.014762 | 0.019905 |
| Flavobacterium | 0.01381 | 0.061031 |
| Anaerobiospirillum | 0.013333 | 0.05209 |
| f__uncultured_Chloroflexi_bacterium | 0.012857 | 0.007838 |
| Izhakiella | 0.012381 | 0.031608 |
| Sporosarcina | 0.011905 | 0.015368 |
| JSC-12 | 0.012381 | 0.0148 |
| Lysobacter | 0.012381 | 0.012611 |
| Rubritepida | 0.010476 | 0.009735 |
| Rubrivirga | 0.01 | 0.01 |
| Actinocatenispora | 0.010476 | 0.048008 |
| metagenome | 0.009524 | 0.010713 |
| Larkinella | 0.009524 | 0.009207 |
| Ruminococcaceae_UCG-014 | 0.009524 | 0.030574 |
| f__Coriobacteriales_Incertae_Sedis_ | 0.008095 | 0.026762 |
| Wilmottia_Ant-Ph58 | 0.008095 | 0.017782 |
| Acidiphilium | 0.009524 | 0.009735 |
| Johnsonella | 0.008095 | 0.030433 |
| JGI_0001001-H03 | 0.00619 | 0.00669 |
| Phascolarctobacterium | 0.006667 | 0.011972 |
| Limnobacter | 0.006667 | 0.009661 |
| Candidatus_Chloroploca | 0.00619 | 0.015961 |
| Rudanella | 0.005714 | 0.010757 |
| f__Acetobacteraceae_Unclassified | 0.005714 | 0.008701 |
| f__uncultured_soil_bacterium_Unclas | 0.004762 | 0.009284 |
| Jan-59 | 0.004762 | 0.006016 |
| f__Family_XIII_Unclassified | 0.005238 | 0.021822 |
| Alloiococcus | 0.004762 | 0.014359 |
| DTU089 | 0.004762 | 0.008136 |
| Nubsella | 0.005238 | 0.006016 |
| Garicola | 0.00381 | 0.01244 |
| Cupriavidus | 0.003333 | 0.006583 |
| Labrys | 0.002857 | 0.007171 |
| Candidatus_Solibacter | 0.002857 | 0.004629 |
| f__Leptolyngbyaceae_Unclassified | 0.003333 | 0.005774 |
| Xylochloris_irregularis | 0.002857 | 0.004629 |
| Verticia | 0.002857 | 0.004629 |
| F0332 | 0.002857 | 0.011019 |
| Corynebacterium | 0.002857 | 0.005606 |
| Myroides | 0.002381 | 0.007003 |
| Luedemannella | 0.002857 | 0.006437 |
| Xanthobacter | 0.002381 | 0.004364 |
| Murinocardiopsis | 0.002857 | 0.009024 |
| Chromobacterium | 0.002381 | 0.010911 |
| Rickettsiella | 0.002381 | 0.00539 |
| FFCH7168 | 0.001905 | 0.004024 |
| Oribacterium | 0.001429 | 0.003586 |
| Treponema | 0.002381 | 0.007003 |
| f__Sandaracinaceae_Unclassified | 0.001905 | 0.006796 |
| Bulleidia | 0.001905 | 0.008729 |
| Planktothrix_NIVA-CYA_15 | 0.001905 | 0.006796 |
| Algoriphagus | 0.001429 | 0.003586 |
| f__P30B-42_Unclassified | 0.001429 | 0.003586 |
| f__Chitinophagaceae_Unclassified | 0.000476 | 0.002182 |
| Pleurocapsa_PCC-7319 | 0.000476 | 0.002182 |
| Alloprevotella | 0.000952 | 0.004364 |
| Longimicrobium | 0.000952 | 0.004364 |
| Chitinimonas | 0.000952 | 0.004364 |
| f__AKIW781_Unclassified | 0.001429 | 0.004781 |
| Collinsella | 0.001429 | 0.003586 |
| Papillibacter | 0.000952 | 0.004364 |
| Christensenellaceae_R-7_group | 0.000476 | 0.002182 |
| f__Deinococcaceae_Unclassified | 0.000952 | 0.003008 |
| f__Geminicoccaceae_Unclassified | 0.000476 | 0.002182 |
| f__Paracaedibacteraceae_Unclassifie | 0.000476 | 0.002182 |
| Chloronema | 0.000476 | 0.002182 |
| f__Erysipelotrichaceae_Unclassified | 0.000476 | 0.002182 |
| | 69.067617 | |

TableS3 Fungi phylum abundance.

| Taxon | average abundance | standard deviation |
| --- | --- | --- |
| Ascomycota | 72.46238 | 11.63144 |
| Basidiomycota | 17.76048 | 8.932234 |
| Unclassified | 8.815238 | 10.52488 |
| k__Fungi_Unclassi | 0.56619 | 0.295643 |
| Zygomycota | 0.350476 | 0.339521 |
| Chytridiomycota | 0.04381 | 0.198481 |
| Glomeromycota | 0.000476 | 0.002182 |
| | 9.77714 | |

TableS4 Fungi genus abundance.

| Taxon | average abundance | standard deviation |
|---|---|---|
| Unclassified | 23.48952 | 10.86516 |
| Aspergillus | 16.59381 | 7.938729 |
| Penicillium | 10.15476 | 15.70518 |
| Hortaea | 7.98 | 7.510961 |
| Cladosporium | 7.811429 | 8.167595 |
| Wallemia | 6.569048 | 9.166376 |
| Emericella | 4.059048 | 9.633092 |
| f__unidentified_Unclass | 3.18381 | 1.156601 |
| Penidiella | 3.02381 | 4.149829 |
| Khuskia | 1.647619 | 1.178541 |
| Candida | 1.256667 | 0.841406 |
| Grammothele | 0.598095 | 0.376757 |
| Sagenomella | 0.540476 | 0.565955 |
| Nigrospora | 0.488095 | 0.390572 |
| Schizophyllum | 0.411429 | 0.19171 |
| Bipolaris | 0.377619 | 0.416808 |
| Eurotium | 0.362381 | 0.576731 |
| Mycosphaerella | 0.342857 | 0.230242 |
| Fusarium | 0.339048 | 0.457569 |
| Coprinellus | 0.328571 | 0.558957 |
| Trichosporon | 0.309524 | 0.19916 |
| Resinicium | 0.30381 | 0.175171 |
| Corynespora | 0.302381 | 0.198038 |
| Letendraea | 0.291429 | 0.095827 |
| Trichomonascus | 0.277619 | 1.269921 |
| Kodamaea | 0.251429 | 0.380398 |
| Choanephora | 0.240952 | 0.317976 |
| Phellinus | 0.238095 | 0.279224 |
| Devriesia | 0.23619 | 0.529136 |
| Aureobasidium | 0.214286 | 0.123068 |
| Gloeotinia | 0.209048 | 0.44579 |
| Auricularia | 0.208571 | 0.118671 |
| Lentinus | 0.187143 | 0.206571 |
| Phoma | 0.177619 | 0.425193 |
| Cerrena | 0.176667 | 0.128154 |
| Sympodiomycopsis | 0.172857 | 0.254463 |
| Hyphodontia | 0.169048 | 0.103533 |
| Coprinopsis | 0.168095 | 0.188563 |
| Leptosphaerulina | 0.162381 | 0.133825 |
| Lasiodiplodia | 0.162381 | 0.125893 |
| Ganoderma | 0.161905 | 0.100081 |
| Curvularia | 0.158095 | 0.140236 |
| Guignardia | 0.151905 | 0.076264 |
| Exidia | 0.14381 | 0.112003 |
| Leptosphaeria | 0.140476 | 0.091677 |
| Cryptococcus | 0.14 | 0.114586 |
| Phlebiopsis | 0.136667 | 0.109011 |
| Neurospora | 0.132381 | 0.083 |
| Ceratobasidium | 0.132857 | 0.404093 |
| Phanerochaete | 0.13381 | 0.102395 |
| Pestalotiopsis | 0.126667 | 0.061427 |
| Eutypa | 0.120952 | 0.055128 |
| Marasmiellus | 0.118571 | 0.094831 |
| f__Corticiaceae_Unclass | 0.116667 | 0.092808 |
| Termitomyces | 0.114286 | 0.076588 |
| Phaeosphaeriopsis | 0.109524 | 0.0574 |
| Peniophora | 0.108571 | 0.06552 |
| Endocarpon | 0.109048 | 0.122593 |
| Malassezia | 0.103333 | 0.080146 |
| f__Herpotrichiellaceae_ | 0.101905 | 0.115309 |
| Sarcinomyces | 0.1 | 0.13896 |
| Pycnoporus | 0.090952 | 0.113574 |
| Saccharomyces | 0.092381 | 0.200198 |
| Limonomyces | 0.088571 | 0.104847 |
| Pleurotus | 0.082857 | 0.057805 |
| Periconia | 0.078571 | 0.122568 |
| f__Trichocomaceae_Uncla | 0.071905 | 0.148142 |
| Diaporthe | 0.070476 | 0.110067 |
| Myrothecium | 0.069048 | 0.063946 |
| Pseudozyma | 0.070476 | 0.080838 |
| Cochliobolus | 0.069048 | 0.069419 |
| Trechispora | 0.067143 | 0.174475 |
| Hyphoderma | 0.06619 | 0.043414 |
| Rigidoporus | 0.06619 | 0.059117 |
| Curreya | 0.06381 | 0.093031 |
| Megasporoporia | 0.052857 | 0.031803 |
| Clavispora | 0.050952 | 0.038458 |
| Gibberella | 0.044762 | 0.026762 |
| Alternaria | 0.045238 | 0.044679 |
| Rhizophlyctis | 0.04381 | 0.198481 |
| Acremonium | 0.039524 | 0.042482 |
| Gymnopilus | 0.038571 | 0.039533 |
| Meyerozyma | 0.040476 | 0.039303 |
| Rhodotorula | 0.038571 | 0.023934 |
| Tritirachium | 0.038095 | 0.047499 |
| Psathyrella | 0.038095 | 0.025811 |
| Debaryomyces | 0.03619 | 0.091295 |
| Stagonospora | 0.034762 | 0.035863 |
| Trametes | 0.03619 | 0.03892 |
| Nakaseomyces | 0.033333 | 0.152753 |
| Cryptodiscus | 0.032857 | 0.037168 |
| Dinemasporium | 0.03381 | 0.045987 |
| f__Russulaceae_Unclassi | 0.03381 | 0.025783 |
| Phialophora | 0.032381 | 0.09375 |
| Dokmaia | 0.030952 | 0.021191 |
| Yamadazyma | 0.030476 | 0.038791 |
| Pilidiella | 0.030952 | 0.023217 |
| Blakeslea | 0.029524 | 0.047799 |
| f__Pleosporaceae_Unclas | 0.027143 | 0.021481 |
| Microsphaeropsis | 0.02619 | 0.019869 |
| Basidiobolus | 0.024762 | 0.048951 |
| Waitea | 0.025238 | 0.047288 |
| Trichothecium | 0.025238 | 0.01721 |
| Amphilogia | 0.024762 | 0.017782 |
| Cercospora | 0.024762 | 0.019905 |
| Marasmius | 0.02381 | 0.020119 |
| Trichaptum | 0.022381 | 0.017862 |
| Lopharia | 0.021905 | 0.016619 |
| Pluteus | 0.020952 | 0.022339 |
| Meira | 0.021905 | 0.023371 |
| Colletotrichum | 0.022381 | 0.011792 |
| Eutypella | 0.021905 | 0.012891 |
| Laetisaria | 0.020476 | 0.066819 |
| Coriolopsis | 0.02 | 0.022804 |
| Zasmidium | 0.019524 | 0.064612 |
| Hypochnicium | 0.018095 | 0.013645 |
| Xylaria | 0.018571 | 0.020563 |
| Microdochium | 0.018095 | 0.024004 |
| Polychaeton | 0.017619 | 0.031923 |
| Physalacria | 0.017619 | 0.009952 |
| Daldinia | 0.016667 | 0.02331 |
| Fusidium | 0.01619 | 0.032936 |
| Plectosphaerella | 0.01619 | 0.013956 |
| Psilocybe | 0.01619 | 0.014992 |
| Phlebia | 0.014762 | 0.013274 |
| Oudemansiella | 0.015714 | 0.01165 |
| f__Botryobasidiaceae_Ur | 0.015714 | 0.01399 |
| Trichoderma | 0.015238 | 0.018873 |
| Leptoxyphium | 0.014762 | 0.022499 |
| Verrucaria | 0.01381 | 0.04153 |
| Orbilia | 0.013333 | 0.024152 |
| Parasympodiella | 0.014286 | 0.013628 |
| Lophiostoma | 0.01381 | 0.011609 |
| Mucor | 0.012857 | 0.054511 |
| Corticium | 0.013333 | 0.011972 |
| Flavodon | 0.01381 | 0.014992 |
| Ramichloridium | 0.012857 | 0.017647 |
| Sistotremastrum | 0.012857 | 0.011464 |
| Hypoxylon | 0.011905 | 0.018335 |
| Sporobolomyces | 0.012381 | 0.013749 |
| Paraphaeosphaeria | 0.012381 | 0.012209 |
| Tetraplosphaeria | 0.011429 | 0.011952 |
| f__Leptosphaeriaceae_Ur | 0.011905 | 0.029261 |
| Panus | 0.011429 | 0.024756 |
| Eremascus | 0.011905 | 0.040696 |
| Blastobotrys | 0.011429 | 0.050128 |
| Microporus | 0.012381 | 0.014108 |
| Hymenochaete | 0.011429 | 0.011526 |
| Hypholoma | 0.010476 | 0.022688 |
| Camillea | 0.011429 | 0.012364 |
| Exidiopsis | 0.01 | 0.008944 |
| Geosmithia | 0.01 | 0.045826 |
| Truncatella | 0.01 | 0.043589 |
| Phaeococcus | 0.01 | 0.017889 |
| Poitrasia | 0.010476 | 0.023765 |
| Gerronema | 0.01 | 0.008944 |
| Capnodium | 0.01 | 0.027749 |
| Hannaella | 0.008571 | 0.014928 |
| Dirinaria | 0.009048 | 0.018683 |
| Nigroporus | 0.009524 | 0.009735 |
| Paecilomyces | 0.009048 | 0.013381 |
| Sydowia | 0.009048 | 0.030316 |
| Valsa | 0.008095 | 0.009284 |
| Dendryphiella | 0.008571 | 0.023725 |
| Parasola | 0.007619 | 0.009437 |
| Ochrocladosporium | 0.008095 | 0.037097 |
| Botryosphaeria | 0.008095 | 0.008729 |
| Inonotus | 0.007143 | 0.009024 |
| Kazachstania | 0.007143 | 0.013093 |
| f__Elsinoaceae_Unclassi | 0.008095 | 0.008136 |
| Passalora | 0.008095 | 0.012891 |
| Phialemonium | 0.007619 | 0.019469 |
| Pyrenochaeta | 0.008095 | 0.010305 |
| Preussia | 0.007143 | 0.010071 |
| Xenasmatella | 0.007143 | 0.011019 |
| Sporisorium | 0.007143 | 0.016169 |
| Humicola | 0.007619 | 0.015781 |
| Monographella | 0.007143 | 0.006437 |
| f__Polyporaceae_Unclass | 0.006667 | 0.007958 |
| Coniothyrium | 0.006667 | 0.011972 |
| Chaetomium | 0.006667 | 0.009129 |
| Eichleriella | 0.008095 | 0.008136 |
| Myrmecridium | 0.005714 | 0.013256 |
| Thielaviopsis | 0.006667 | 0.009129 |
| Arthrobotrys | 0.005714 | 0.026186 |
| Polyporus | 0.00619 | 0.00669 |
| Phaeosaccardinula | 0.005238 | 0.014359 |
| Sterigmatomyces | 0.004762 | 0.009284 |
| Vascellum | 0.006667 | 0.006583 |
| Sporidiobolus | 0.004762 | 0.009808 |
| Magnaporthe | 0.005714 | 0.007464 |
| Glomerella | 0.005238 | 0.011233 |
| Gibellulopsis | 0.004762 | 0.013645 |
| Dissoconium | 0.005714 | 0.006761 |
| Meripilus | 0.005714 | 0.008106 |
| Brycekendrickomyces | 0.005714 | 0.010282 |
| Perenniporia | 0.005238 | 0.011233 |
| Phaeosphaeria | 0.005238 | 0.01504 |
| Bullera | 0.005238 | 0.006016 |
| Annulohypoxylon | 0.005714 | 0.007464 |
| f__Microbotryaceae_Uncl | 0.005238 | 0.006016 |
| f__Massarinaceae_Unclas | 0.00381 | 0.005896 |
| Echinochaete | 0.004762 | 0.009808 |
| Pichia | 0.004286 | 0.008106 |
| Stilbohypoxylon | 0.004762 | 0.007496 |
| Moesziomyces | 0.005238 | 0.006796 |
| f__Arthopyreniaceae_Unc | 0.004762 | 0.006016 |
| Chaetomella | 0.005238 | 0.008136 |
| Saccharomycopsis | 0.004286 | 0.01964 |
| Tilletiopsis | 0.004762 | 0.006016 |
| Stenella | 0.004286 | 0.006761 |
| Emericellopsis | 0.00381 | 0.009735 |
| Biscogniauxia | 0.003333 | 0.013166 |
| Fomitopsis | 0.00381 | 0.004976 |
| Volutella | 0.003333 | 0.013166 |
| f__Phaeosphaeriaceae_Ur | 0.00381 | 0.005896 |
| Torulaspora | 0.003333 | 0.00483 |
| Mortierella | 0.003333 | 0.015275 |
| Ceriporiopsis | 0.003333 | 0.009129 |
| Aplosporella | 0.003333 | 0.011547 |
| Isaria | 0.002381 | 0.00539 |
| f__Bionectriaceae_Uncla | 0.003333 | 0.00483 |
| Camarosporium | 0.002857 | 0.004629 |
| Rhizopus | 0.002857 | 0.004629 |
| Gymnopus | 0.002857 | 0.004629 |
| Calyptella | 0.002857 | 0.009562 |
| Rhodosporidium | 0.002381 | 0.010911 |
| Microascus | 0.002857 | 0.009024 |
| Podospora | 0.002857 | 0.009024 |
| Lodderomyces | 0.001429 | 0.003586 |
| Harpophora | 0.001905 | 0.004024 |
| Septobasidium | 0.001905 | 0.008729 |
| Xenostigmina | 0.001905 | 0.006016 |
| Tremella | 0.001905 | 0.008729 |
| Zygoascus | 0.001429 | 0.006547 |
| Sporothrix | 0.001905 | 0.006796 |
| Hanseniaspora | 0.001905 | 0.006796 |
| Berkleasmium | 0.001429 | 0.003586 |
| Ascobolus | 0.001429 | 0.004781 |
| Byssochlamys | 0.001429 | 0.004781 |
| Resupinatus | 0.000952 | 0.003008 |
| Hohenbuehelia | 0.001429 | 0.004781 |
| Rhizomucor | 0.001429 | 0.003586 |
| f__Amphisphaeriaceae_Ur | 0.000952 | 0.003008 |
| f__Valsaceae_Unclassifi | 0.000952 | 0.004364 |
| Infundibulomyces | 0.000476 | 0.002182 |
| Cryptotrichosporon | 0.000952 | 0.003008 |
| f__Agaricaceae_Unclassi | 0.000476 | 0.002182 |
| Neodeightonia | 0.000952 | 0.003008 |
| f__Ascobolaceae_Unclass | 0.000952 | 0.004364 |
| f__Cortinariaceae_Uncla | 0.000952 | 0.003008 |
| Acanthostigma | 0.000476 | 0.002182 |
| Cladonia | 0.000476 | 0.002182 |
| Hemibeltrania | 0.000476 | 0.002182 |
| Teratosphaeria | 0.000476 | 0.002182 |
| | 46.831905 | |

(alpha diversity). P value was calculated based on 10,000 permutation bivariate Adonis analysis.

| | | bacteria | | | fungi | | |
|---|---|---|---|---|---|---|---|
| | | Median (p25-p75) | $R^2$ | p | Median (p25-p75) | $R^2$ | p |
| Relative humidity | | | 0.04 | 0.39 | | 0.03 | 0.48 |
| | low | 678.95 (647.3-682.7) | | | 672.85 (625.1-736.175) | | |
| | high | 691.5 (615.5-700.4) | | | 783.9 (560.6-853.25) | | |
| CO2 indoor | | | 0.02 | 0.6 | | 0.07 | 0.25 |
| | low | 679.2 (647.3-685.8) | | | 659 (618.875-712.125) | | |
| | high | 683.3 (667.1-700.4) | | | 806 (776.25-853.25) | | |
| NO2 outdoor | | | 0.1 | 0.16 | | 0.001 | 0.87 |
| | low | 676.9 （645.3-686.7） | | | 662.3 (614.9-773.3） | | |
| | high | 683.2 (640.9-704.575) | | | 781.5 (560.6-830.95) | | |
| Building age | | | 0 | 0.99 | | 0.004 | 0.79 |
| | low | 676.9 （656.8-687.3） | | | 721.7 (643.25-792.6) | | |
| | high | 685 (615.5-696.925） | | | 778.6 (400.7-848.43) | | |
| Dampness | | | 0 | 0.9 | | 0.01 | 0.63 |
| | low | 678.95 （647.3-682.7） | | | 672.85 (625.1-736.18) | | |
| | high | 691.5 (615.5-700.4） | | | 783.9 (560.6-853.25) | | |
| Curtain | | | 0.03 | 0.42 | | **0.33** | **0.006** |
| | low | 682.15<br>(663.7-691.28） | | | 814<br>(774.8-848.43) | | |
| | high | 681.5<br>（638.9-697） | | | 655.7<br>(400.7-731.35) | | |
| House dustmite | | | 0.09 | 0.18 | | **0.28** | **0.02** |
| | low | 689.1<br>(676.25-696.93) | | | 826.8<br>(774.78-855.5) | | |
| | high | 676.9<br>（638.9-682.3） | | | 662.3<br>(400.7-731.35) | | |
| Cockroach | | | 0.002 | 0.84 | | **0.41** | **0.002** |
| | low | 683.3<br>（673.9-698.3） | | | 779.2<br>（761-857.8） | | |
| | high | 678.95<br>(638.9-691.825） | | | 659<br>(400.7-789.43) | | |

TableS6 Association between OTUs and asthma.

| | Odds Ratio | p值 | 95CI | |
|---|---|---|---|---|
| bac_phylum_Actinobacteria_138_otu | 1.016937 | 0.518 | 0.966426 | 1.070087 |
| bac_phylum_Cyanobacteria_156_otu | 0.9771775 | 0.071 | 0.953032 | 1.001935 |
| bac_phylum_Firmicutes_139_otu | 0.9986067 | 0.951 | 0.954921 | 1.044291 |
| bac_phylum_Proteobacteria_242_otu | 0.9639052 | 0.089 | 0.923883 | 1.005661 |
| bac_phylum_Bacteroidetes_95_otu | 1.014582 | 0.408 | 0.980408 | 1.049948 |
| bac_phylum_Deinococcus_Thermus_ | 1.024335 | 0.627 | 0.92964 | 1.128676 |
| bac_phylum_other_125_otu | 0.9939591 | 0.678 | 0.965921 | 1.022811 |
| bac_total_otu_895 | 0.9957413 | 0.411 | 0.98567 | 1.005916 |
| | | | | |
| bac_class__Actinobacteria | 0.9916803 | 0.805 | 0.927885 | 1.059862 |
| bac_class__Alphaproteobacteria | 0.9434536 | 0.122 | 0.876368 | 1.015675 |
| bac_class__Oxyphotobacteria | 0.980741 | 0.16 | 0.954504 | 1.007699 |
| bac_class__Gammaproteobacteria | 0.9944438 | 0.888 | 0.920043 | 1.074861 |
| bac_class__Bacteroidia | 1.016516 | 0.388 | 0.979391 | 1.055048 |
| bac_class__Bacilli | 0.9847966 | 0.724 | 0.904411 | 1.072327 |
| | | | | |
| fu_phylum_Ascomycota_763_otu | 0.9992615 | 0.748 | 0.994768 | 1.003775 |
| fu_phylum_Basidiomycota_453_otu | 0.9987332 | 0.681 | 0.992713 | 1.00479 |
| fu_phylum_unidentified_36 | 0.946967 | 0.127 | 0.882949 | 1.015626 |
| fu_phylum_other_260 | 0.9995034 | 0.941 | 0.986452 | 1.012728 |
| fu_total_otu_1512 | 0.9996034 | 0.72 | 0.997438 | 1.001773 |
| | | | | |
| fu_class__Agaricomycetes | 0.9984552 | 0.734 | 0.98958 | 1.00741 |
| fu_class__Eurotiomycetes | 1.008482 | 0.547 | 0.981139 | 1.036587 |
| fu_class__Dothideomycetes | 1.000539 | 0.93 | 0.988527 | 1.012697 |
| fu_class__Sordariomycetes | 0.9862998 | 0.215 | 0.96502 | 1.008049 |
| fu_class__Tremellomycetes | 0.9988693 | 0.962 | 0.953175 | 1.046754 |
| fu_class__Saccharomycetes | 0.9841009 | 0.624 | 0.923101 | 1.049132 |

TableS7 Bacterial absolate abundance vs relative abundance.

| | | Relative abundance | Absolute abundance |
|---|---|---|---|
| RAbac1 | Unclassified | 0.516 | 0.715 |
| RAbac10 | f__Chroococcidiopsaceae_Unc | 0.578 | 0.386 |
| RAbac100 | Terriglobus | 0.132 | 0.91 |
| RAbac101 | Dolosigranulum | 0.332 | 0.78 |
| RAbac102 | f__uncultured_bacterium_Unc | 0.664 | 0.605 |
| RAbac103 | Amaricoccus | 0.183 | 0.044 |
| RAbac104 | Leptolyngbya_ANT.L52.2 | 0.487 | 0.243 |
| RAbac105 | f__Roseiflexaceae_Unclassif | 0.648 | 0.929 |
| RAbac106 | Blautia | 0.212 | 0.282 |
| RAbac107 | f__Caulobacteraceae_Unclass | 0.5 | 0.699 |
| RAbac108 | Chalicogloea_CCALA_975 | 0.199 | 0.091 |
| RAbac109 | Fusobacterium | 0.223 | 0.048 |
| RAbac11 | Corynebacterium_1 | 0.964 | 0.225 |
| RAbac110 | Microvirga | 0.01 | 0.075 |
| RAbac111 | Ornithinimicrobium | 0.776 | 0.798 |
| RAbac112 | RB41 | 0.804 | 0.904 |
| RAbac113 | Cellulomonas | 0.832 | 0.921 |
| RAbac114 | Alkanindiges | 0.903 | 0.556 |
| RAbac115 | Dietzia | 0.765 | 0.921 |
| RAbac116 | Pseudocitrobacter | 0.822 | 0.55 |
| RAbac117 | Piscicoccus | 0.406 | 0.642 |
| RAbac118 | Kineosporia | 0.474 | 0.962 |
| RAbac119 | Halomonas | 0.354 | 0.508 |
| RAbac12 | Bacteroides | 0.852 | 0.72 |
| RAbac120 | Sphingobium | 0.142 | 0.004 |
| RAbac121 | Aureimonas | 0.596 | 0.415 |
| RAbac122 | Comamonas | 0.688 | 0.412 |
| RAbac123 | Dermacoccus | 0.086 | 0.122 |
| RAbac124 | Sandaracinobacter | 0.578 | 0.912 |
| RAbac125 | Kineococcus | 0.483 | 0.555 |
| RAbac126 | Streptomyces | 0.838 | 0.79 |
| RAbac127 | Bryocella | 0.574 | 0.913 |
| RAbac128 | Pleurocapsa_PCC-7327 | 0.002 | 0.004 |
| RAbac129 | Rheinheimera | 0.949 | 0.367 |
| RAbac13 | Rubellimicrobium | 0.482 | 0.997 |
| RAbac130 | Flavisolibacter | 0.158 | 0.2 |
| RAbac131 | Fictibacillus | 0.677 | 0.472 |
| RAbac132 | Altererythrobacter | 0.461 | 0.548 |
| RAbac133 | Aerococcus | 0.41 | 0.21 |
| RAbac134 | Sphingobacterium | 0.097 | 0.867 |
| RAbac135 | Belnapia | 0.496 | 0.467 |
| RAbac136 | Shewanella | 0.406 | 0.929 |
| RAbac137 | Achromobacter | 0.366 | 0.818 |
| RAbac138 | Devosia | 0.205 | 0.24 |
| RAbac139 | Finegoldia | 0.489 | 0.342 |
| RAbac14 | Pantoea | 0.995 | 0.916 |
| RAbac140 | Nakamurella | 0.817 | 0.362 |
| RAbac141 | Arthrobacter | 0.503 | 0.704 |
| RAbac142 | Mastigocladopsis_PCC-10914 | 0.429 | 0.109 |
| RAbac143 | Scytonema_PCC-7110 | 0.509 | 0.16 |
| RAbac144 | Serratia | 0.719 | 0.083 |
| RAbac145 | endosymbionts8 | 0.487 | 0.368 |
| RAbac146 | Megamonas | 0.658 | 0.634 |
| RAbac147 | Hungatella | 0.025 | 0.446 |
| RAbac148 | Salinicoccus | 0.528 | 0.621 |
| RAbac149 | Wolbachia | 0.005 | 0.762 |
| RAbac15 | Streptococcus | 0.129 | 0.085 |
| RAbac150 | Prevotella_7 | 0.591 | 0.57 |
| RAbac151 | Cnuella | 0.134 | 0.319 |
| RAbac152 | Granulicatella | 0.171 | 0.032 |
| RAbac153 | Aliicoccus | 0.347 | 0.972 |
| RAbac154 | Candidatus_Alysiosphaera | 0.864 | 0.288 |
| RAbac155 | Gemella | 0.881 | 0.04 |
| RAbac156 | Sphingoaurantiacus | 0.909 | 0.454 |
| RAbac157 | Pseudoxanthomonas | 0.451 | 0.972 |
| RAbac158 | Antricoccus | 0.732 | 0.413 |
| RAbac159 | Scytonema_UTEX_2349 | 0.316 | 0.416 |
| RAbac16 | Kocuria | 0.574 | 0.836 |
| RAbac160 | Veillonella | 0.363 | 0.432 |
| RAbac161 | Paeniclostridium | 0.155 | 0.946 |
| RAbac162 | Lysinibacillus | 0.535 | 0.994 |
| RAbac163 | f__Coleofasciculaceae_Uncla | 0.164 | 0.658 |
| RAbac164 | f__Archangiaceae_Unclassifi | 0.147 | 0.064 |
| RAbac165 | Actinoplanes | 0.709 | 0.269 |
| RAbac166 | Vittaria_lineata__[shoestri | 0.083 | 0.11 |
| RAbac167 | Defluviicoccus | 0.171 | 0.18 |
| RAbac168 | Lawsonella | 0.631 | 0.259 |
| RAbac169 | Leptolyngbya_VRUC_135 | 0.919 | 0.872 |
| RAbac17 | Brevibacterium | 0.438 | 0.393 |
| RAbac170 | Roseburia | 0.24 | 0.829 |
| RAbac171 | Peptoniphilus | 0.263 | 0.911 |
| RAbac172 | f__Ambiguous_taxa_Unclassif | 0.786 | 0.761 |
| RAbac173 | Cloacibacterium | 0.677 | 0.582 |
| RAbac174 | Solirubrobacter | 0.057 | 0.003 |
| RAbac175 | Odoribacter | 0.313 | 0.381 |
| RAbac176 | Catellicoccus | 0.113 | 0.832 |
| RAbac177 | Marinococcus | 0.609 | 0.283 |
| RAbac178 | Faecalibacterium | 0.661 | 0.709 |
| RAbac179 | Lautropia | 0.505 | 0.029 |
| RAbac18 | Methylobacterium | 0.456 | 0.835 |
| RAbac180 | Eggerthella | 0.095 | 0.174 |
| RAbac181 | Aetokthonos_AEL04 | 0.788 | 0.902 |
| RAbac182 | Paenibacillus | 0.113 | 0.454 |
| RAbac183 | Luteimonas | 0.805 | 0.828 |
| RAbac184 | Tatumella | 0.946 | 0.422 |
| RAbac185 | Porphyrobacter | 0.659 | 0.036 |
| RAbac186 | f__uncultured_Chloroflexi_b | 0.176 | 0.34 |
| RAbac187 | f__Saccharimonadaceae_Uncla | 0.595 | 0.529 |
| RAbac188 | Peptoclostridium | 0.275 | 0.695 |
| RAbac189 | Porphyromonas | 0.378 | 0.826 |
| RAbac19 | Rubrobacter | 0.43 | 0.453 |
| RAbac190 | Akkermansia | 0.642 | 0.062 |
| RAbac191 | Bifidobacterium | 0.607 | 0.212 |
| RAbac192 | Rhodomicrobium | 0.942 | 0.002 |
| RAbac193 | Granulicella | 0.526 | 0.574 |
| RAbac194 | Proteus | 0.315 | 0.766 |
| RAbac195 | f__Corynebacteriaceae_Uncla | 0.668 | 0.924 |
| RAbac196 | Providencia | 0.144 | 0.283 |
| RAbac197 | f__Lachnospiraceae_Unclassi | 0.833 | 0.092 |
| RAbac198 | Peptostreptococcus | 0.195 | 0.361 |
| RAbac199 | [Ruminococcus]_torques_grou | 0.434 | 0.212 |
| RAbac2 | Bacillus | 0.418 | 0.407 |
| RAbac20 | Haemophilus | 0.318 | 0.04 |
| RAbac200 | Saccharibacillus | 0.488 | 0.825 |
| RAbac201 | Eubacterium | 0.55 | 0.57 |
| RAbac202 | f__Xanthobacteraceae_Unclas | 0.486 | 0.986 |
| RAbac203 | Thauera | 0.01 | 0.615 |
| RAbac204 | YB-42 | 0.875 | 0.054 |
| RAbac205 | EcFYyy-200 | 0.492 | 0.333 |
| RAbac206 | Moraxella | 0.247 | 0.105 |
| RAbac207 | Craurococcus | 0.266 | 0.072 |
| RAbac208 | [Ruminococcus]_gnavus_group | 0.178 | 0.277 |
| RAbac209 | Microcoleus_SAG_1449-1a | 0.653 | 0.014 |
| RAbac21 | Pseudonocardia | 0.853 | 0.929 |
| RAbac210 | Nocardiopsis | 0.001 | 0.276 |
| RAbac211 | Cedecea | 0.864 | 0.165 |
| RAbac212 | [Eubacterium]_fissicatena_g | 0.681 | 0.522 |
| RAbac213 | Blastocatella | 0.218 | 0.275 |
| RAbac214 | Ochrobactrum | 0.614 | 0.546 |
| RAbac215 | Mesorhizobium | 0.641 | 0.521 |
| RAbac216 | Lysobacter | 0.795 | 0.95 |
| RAbac217 | Rubritepida | 0.231 | 0.082 |
| RAbac218 | Larkinella | 0.091 | 0.097 |
| RAbac219 | Gardnerella | 0.088 | 0.119 |
| RAbac22 | Deinococcus | 0.211 | 0.753 |
| RAbac220 | Gemmatirosa | 0.055 | 0.156 |
| RAbac221 | f__Caldilineaceae_Unclassif | 0.866 | 0.607 |
| RAbac222 | [Eubacterium]_coprostanolig | 0.679 | 0.855 |
| RAbac223 | Terrisporobacter | 0.178 | 0.252 |
| RAbac224 | Rubrivirga | 0.621 | 0.28 |
| RAbac225 | Bounagaea | 0.541 | 0.032 |
| RAbac226 | Leptotrichia | 0.39 | 0.73 |
| RAbac227 | f__Nostocaceae_Unclassified | 0.674 | 0.591 |
| RAbac228 | Vibrio | 0.349 | 0.322 |
| RAbac229 | Acidothermus | 0.615 | 0.131 |
| RAbac23 | Chroococcidiopsis_SAG_2023 | 0.913 | 0.806 |
| RAbac230 | Microseira_Carmichael-Alaba | 0.674 | 0.735 |
| RAbac231 | Robinsoniella | 0.001 | 0.004 |
| RAbac232 | Acidiphilium | 0.047 | 0.315 |
| RAbac233 | Haliangium | 0.21 | 0.171 |
| RAbac234 | Aggregatibacter | 0.578 | 0.726 |
| RAbac235 | Adhaeribacter | 0.219 | 0.138 |
| RAbac236 | Abiotrophia | 0.54 | 0.075 |
| RAbac237 | f__A4b_Unclassified | 0.826 | 0.795 |
| RAbac238 | Megasphaera | 0.132 | 0.41 |
| RAbac239 | Bilophila | 0.426 | 0.045 |
| RAbac24 | Enterococcus | 0.426 | 0.206 |
| RAbac240 | Anaerostipes | 0.668 | 0.676 |
| RAbac241 | Alistipes | 0.47 | 0.1 |
| RAbac242 | JSC-12 | 0.589 | 0.341 |
| RAbac243 | metagenome | 0.775 | 0.275 |
| RAbac244 | Capnocytophaga | 0.381 | 0.409 |
| RAbac245 | Prauserella | 0.381 | 0.731 |
| RAbac246 | Rikenella | 0.027 | 0.07 |
| RAbac247 | Microcoleus_Es-Yyy1400 | 0.289 | 0.236 |
| RAbac248 | Cyanothece_PCC-7424 | 0.236 | 0.339 |
| RAbac249 | Vagococcus | 0.889 | 0.855 |
| RAbac25 | Micrococcus | 0.707 | 0.582 |
| RAbac250 | Solibacillus | 0.631 | 0.181 |
| RAbac251 | GCA-900066225 | 0.941 | 0.95 |
| RAbac252 | Sporosarcina | 0.235 | 0.772 |
| RAbac253 | JGI_0001001-H03 | 0.048 | 0.003 |
| RAbac254 | Chakia_8 | 0.831 | 0.77 |
| RAbac255 | Listeria | 0.761 | 0.16 |
| RAbac256 | f__Rhizobiaceae_Unclassifie | 0.623 | 0.407 |
| RAbac257 | Nubsella | 0.883 | 0.428 |
| RAbac258 | Prevotella_2 | 0.524 | 0.762 |
| RAbac259 | Wohlfahrtiimonas | 0.093 | 0.971 |
| RAbac26 | Calothrix_PCC-6303 | 0.84 | 0.547 |
| RAbac260 | Limnobacter | 0.342 | 0.614 |
| RAbac261 | 21551 | 0.135 | 0.207 |
| RAbac262 | Marinilactibacillus | 0.834 | 0.158 |
| RAbac263 | Turicella | 0.897 | 0.805 |
| RAbac264 | Mucilaginibacter | 0.15 | 0.126 |
| RAbac265 | f__Acetobacteraceae_Unclass | 0.658 | 0.395 |
| RAbac266 | Ignatzschineria | 0.402 | 0.991 |
| RAbac267 | f__Euzebyaceae_Unclassified | 0.803 | 0.583 |
| RAbac268 | Phascolarctobacterium | 0.794 | 0.989 |
| RAbac269 | Rudanella | 0.702 | 0.452 |
| RAbac27 | Pseudomonas | 0.643 | 0.366 |
| RAbac270 | f__uncultured_soil_bacteriu | 0.636 | 0.953 |
| RAbac271 | DTU089 | 0.075 | 0.472 |
| RAbac272 | Campylobacter | 0.502 | 0.12 |
| RAbac273 | Plesiomonas | 0.218 | 0.798 |
| RAbac274 | Wilmottia_Ant-Ph58 | 0.531 | 0.518 |
| RAbac275 | Candidatus_Solibacter | 0.153 | 0.158 |
| RAbac276 | f__Leptolyngbyaceae_Unclass | 0.687 | 0.827 |
| RAbac277 | Xylochloris_irregularis | 0.774 | 0.766 |
| RAbac278 | Verticia | 0.425 | 0.398 |
| RAbac279 | Anaerobiospirillum | 0.201 | 0.433 |
| RAbac28 | Massilia | 0.055 | 0.129 |
| RAbac280 | Izhakiella | 0.103 | 0.006 |
| RAbac281 | Candidatus_Chloroploca | 0.773 | 0.817 |
| RAbac282 | Cupriavidus | 0.924 | 0.546 |
| RAbac283 | Corynebacterium | 0.219 | 0.072 |
| RAbac284 | Xanthobacter | 0.469 | 0.405 |
| RAbac29 | Klebsiella | 0.239 | 0.788 |
| RAbac3 | Staphylococcus | 0.875 | 0.42 |
| RAbac30 | Nephrolepis_biserrata_var._ | 0.668 | 0.954 |
| RAbac31 | Brachybacterium | 0.742 | 0.553 |
| RAbac32 | Skermanella | 0.614 | 0.904 |
| RAbac33 | Parabacteroides | 0.554 | 0.8 |
| RAbac34 | Janibacter | 0.266 | 0.308 |
| RAbac35 | Roseomonas | 0.11 | 0.576 |
| RAbac36 | Actinomycetospora | 0.582 | 0.865 |
| RAbac37 | Escherichia-Shigella | 0.333 | 0.164 |
| RAbac38 | Truepera | 0.428 | 0.237 |
| RAbac39 | MN_122.2a | 0.88 | 0.565 |
| RAbac4 | Paracoccus | 0.184 | 0.129 |
| RAbac40 | Geodermatophilus | 0.174 | 0.308 |
| RAbac41 | Curtobacterium | 0.305 | 0.35 |
| RAbac42 | Scytonema_UCFS19 | 0.557 | 0.169 |
| RAbac43 | Chroococcidiopsis_PCC_7203 | 0.749 | 0.362 |
| RAbac44 | Nocardioides | 0.384 | 0.291 |
| RAbac45 | Leptolyngbya_PCC-6306 | 0.619 | 0.98 |
| RAbac46 | Nesterenkonia | 0.405 | 0.242 |
| RAbac47 | CENA359 | 0.368 | 0.271 |
| RAbac48 | Marmoricola | 0.62 | 0.272 |
| RAbac49 | Clostridium_sensu_stricto_1 | 0.434 | 0.883 |
| RAbac5 | Sphingomonas | 0.473 | 0.634 |
| RAbac50 | Brevundimonas | 0.358 | 0.196 |
| RAbac51 | Blastococcus | 0.652 | 0.908 |
| RAbac52 | Enhydrobacter | 0.048 | 0.04 |
| RAbac53 | Exiguobacterium | 0.859 | 0.822 |
| RAbac54 | 1174-901-12 | 0.104 | 0.891 |
| RAbac55 | Scytonema_VB-61278 | 0.903 | 0.417 |
| RAbac56 | Gordonia | 0.832 | 0.378 |
| RAbac57 | Hymenobacter | 0.608 | 0.882 |
| RAbac58 | Dapisostemonum_CCIBt_3536 | 0.559 | 0.453 |
| RAbac59 | Lachnoclostridium | 0.808 | 0.517 |
| RAbac6 | Saccharopolyspora | 0.975 | 0.535 |
| RAbac60 | Pseudokineococcus | 0.817 | 0.474 |
| RAbac61 | Allorhizobium-Neorhizobium- | 0.757 | 0.355 |
| RAbac62 | Neisseria | 0.445 | 0.14 |
| RAbac63 | f__uncultured_Unclassified | 0.47 | 0.634 |
| RAbac64 | Weissella | 0.093 | 0.292 |
| RAbac65 | Kurthia | 0.174 | 0.18 |
| RAbac66 | Stenotrophomonas | 0.691 | 0.986 |
| RAbac67 | Mycobacterium | 0.757 | 0.415 |
| RAbac68 | Romboutsia | 0.041 | 0.58 |
| RAbac69 | Glutamicibacter | 0.822 | 0.634 |
| RAbac7 | Acinetobacter | 0.136 | 0.892 |
| RAbac70 | Bryum_argenteum_var._argent | 0.388 | 0.269 |
| RAbac71 | f__uncultured_cyanobacteriu | 0.75 | 0.684 |
| RAbac72 | Rothia | 0.226 | 0.067 |
| RAbac73 | Shimwellia | 0.002 | 0.002 |
| RAbac74 | Quadrisphaera | 0.059 | 0.542 |
| RAbac75 | Aeromonas | 0.255 | 0.191 |
| RAbac76 | Novosphingobium | 0.432 | 0.272 |
| RAbac77 | f__Blastocatellaceae_Unclas | 0.762 | 0.613 |
| RAbac78 | Bryobacter | 0.802 | 0.799 |
| RAbac79 | Desulfovibrio | 0.964 | 0.418 |
| RAbac8 | Enterobacter | 0.247 | 0.994 |
| RAbac80 | Qipengyuania | 0.02 | 0.025 |
| RAbac81 | Spirosoma | 0.641 | 0.661 |
| RAbac82 | Barrientosiimonas | 0.189 | 0.683 |
| RAbac83 | Aliterella_CENA595 | 0.275 | 0.255 |
| RAbac84 | Chryseobacterium | 0.217 | 0.106 |
| RAbac85 | Lactococcus | 0.404 | 0.244 |
| RAbac86 | Cellulosimicrobium | 0.581 | 0.724 |
| RAbac87 | Kosakonia | 0.609 | 0.181 |
| RAbac88 | Actinomyces | 0.526 | 0.153 |
| RAbac89 | Micromonospora | 0.379 | 0.986 |
| RAbac9 | Lactobacillus | 0.438 | 0.756 |
| RAbac90 | Alishewanella | 0.2 | 0.849 |
| RAbac91 | Jeotgalicoccus | 0.108 | 0.251 |
| RAbac92 | f__Ilumatobacteraceae_Uncla | 0.663 | 0.333 |
| RAbac93 | f__Beijerinckiaceae_Unclass | 0.889 | 0.397 |
| RAbac94 | f__Unknown_Family_Unclassif | 0.165 | 0.621 |
| RAbac95 | f__JG30-KF-CM45_Unclassifie | 0.67 | 0.702 |
| RAbac96 | Kytococcus | 0.49 | 0.394 |
| RAbac97 | PMMR1 | 0.537 | 0.985 |
| RAbac98 | Empedobacter | 0.412 | 0.083 |
| RAbac99 | Modestobacter | 0.141 | 0.369 |

TableS8 Fungal absolate abundance vs relative abundance.

| | | Relative abundanc | Absolute abundance |
|---|---|---|---|
| RAfu1 | Unclassified | 0.389 | 0.433 |
| RAfu2 | Aspergillus | 0.749 | 0.877 |
| RAfu3 | Penicillium | 0.264 | 0.564 |
| RAfu4 | Hortaea | 0.074 | 0.627 |
| RAfu5 | Cladosporium | 0.427 | 0.42 |
| RAfu6 | Wallemia | 0.468 | 0.615 |
| RAfu7 | Emericella | 0.52 | 0.229 |
| RAfu8 | f__unidentified_Unclass | 0.22 | 0.437 |
| RAfu9 | Penidiella | 0.129 | 0.557 |
| RAfu10 | Khuskia | 0.741 | 0.546 |
| RAfu11 | Candida | 0.55 | 0.478 |
| RAfu12 | Grammothele | 0.282 | 0.372 |
| RAfu13 | Sagenomella | 0.7 | 0.769 |
| RAfu14 | Nigrospora | 0.593 | 0.512 |
| RAfu15 | Schizophyllum | 0.543 | 0.483 |
| RAfu16 | Bipolaris | 0.078 | 0.152 |
| RAfu17 | Eurotium | 0.043 | 0.868 |
| RAfu18 | Mycosphaerella | 0.691 | 0.603 |
| RAfu19 | Fusarium | 0.431 | 0.163 |
| RAfu20 | Trichosporon | 0.848 | 0.625 |
| RAfu21 | Resinicium | 0.405 | 0.522 |
| RAfu22 | Corynespora | 0.073 | 0.03 |
| RAfu23 | Letendraea | 0.021 | 0.176 |
| RAfu24 | Kodamaea | 0.276 | 0.925 |
| RAfu25 | Phellinus | 0.685 | 0.512 |
| RAfu26 | Aureobasidium | 0.826 | 0.636 |
| RAfu27 | Auricularia | 0.681 | 0.579 |
| RAfu28 | Phoma | 0.045 | 0.071 |
| RAfu29 | Cerrena | 0.213 | 0.433 |
| RAfu30 | Hyphodontia | 0.366 | 0.444 |
| RAfu31 | Coprinopsis | 0.751 | 0.325 |
| RAfu32 | Leptosphaerulina | 0.915 | 0.538 |
| RAfu33 | Lasiodiplodia | 0.53 | 0.783 |
| RAfu34 | Ganoderma | 0.886 | 0.676 |
| RAfu35 | Curvularia | 0.358 | 0.881 |
| RAfu36 | Guignardia | 0.826 | 0.838 |
| RAfu37 | Exidia | 0.283 | 0.278 |
| RAfu38 | Leptosphaeria | 0.817 | 0.637 |
| RAfu39 | Cryptococcus | 0.111 | 0.355 |
| RAfu40 | Phlebiopsis | 0.4 | 0.488 |
| RAfu41 | Neurospora | 0.584 | 0.984 |
| RAfu42 | Pestalotiopsis | 0.44 | 0.509 |
| RAfu43 | Eutypa | 0.799 | 0.732 |
| RAfu44 | Marasmiellus | 0.651 | 0.477 |
| RAfu45 | f__Corticiaceae_Unclass | 0.751 | 0.723 |
| RAfu46 | Termitomyces | 0.739 | 0.849 |
| RAfu47 | Peniophora | 0.762 | 0.688 |
| RAfu48 | Malassezia | 0.286 | 0.434 |
| RAfu49 | Saccharomyces | 0.034 | 0.094 |
| RAfu50 | Pleurotus | 0.552 | 0.266 |
| RAfu51 | Diaporthe | 0.241 | 0.105 |
| RAfu52 | Myrothecium | 0.043 | 0.076 |
| RAfu53 | Pseudozyma | 0.178 | 0.466 |
| RAfu54 | Hyphoderma | 0.487 | 0.413 |
| RAfu55 | Megasporoporia | 0.923 | 0.86 |
| RAfu56 | Gibberella | 0.998 | 0.781 |
| RAfu57 | Alternaria | 0.704 | 0.423 |
| RAfu58 | Psathyrella | 0.043 | 0.237 |
| RAfu59 | Colletotrichum | 0.02 | 0.126 |
| RAfu60 | Coprinellus | 0.468 | 0.756 |
| RAfu61 | Choanephora | 0.938 | 0.979 |
| RAfu62 | Gloeotinia | 0.793 | 0.728 |
| RAfu63 | Lentinus | 0.785 | 0.938 |
| RAfu64 | Sympodiomycopsis | 0.947 | 0.385 |
| RAfu65 | Phanerochaete | 0.574 | 0.908 |
| RAfu66 | Phaeosphaeriopsis | 0.563 | 0.821 |
| RAfu67 | Limonomyces | 0.093 | 0.113 |
| RAfu68 | Periconia | 0.168 | 0.809 |
| RAfu69 | Rhodotorula | 0.571 | 0.952 |
| RAfu70 | Dokmaia | 0.335 | 0.803 |
| RAfu71 | Pilidiella | 0.474 | 0.576 |
| RAfu72 | Devriesia | 0.866 | 0.752 |
| RAfu73 | Endocarpon | 0.337 | 0.732 |
| RAfu74 | f__Herpotrichiellaceae_ | 0.235 | 0.989 |
| RAfu75 | Pycnoporus | 0.807 | 0.617 |
| RAfu76 | Trechispora | 0.73 | 0.716 |
| RAfu77 | Rigidoporus | 0.469 | 0.549 |
| RAfu78 | Curreya | 0.512 | 0.637 |
| RAfu79 | Clavispora | 0.833 | 0.805 |
| RAfu80 | Tritirachium | 0.067 | 0.578 |
| RAfu81 | Trametes | 0.643 | 0.808 |
| RAfu82 | Cryptodiscus | 0.212 | 0.25 |
| RAfu83 | Dinemasporium | 0.955 | 0.767 |
| RAfu84 | f__Russulaceae_Unclassi | 0.439 | 0.691 |
| RAfu85 | Blakeslea | 0.082 | 0.719 |
| RAfu86 | f__Pleosporaceae_Unclas | 0.431 | 0.903 |
| RAfu87 | Amphilogia | 0.051 | 0.319 |
| RAfu88 | Marasmius | 0.197 | 0.267 |
| RAfu89 | Ceratobasidium | 0.581 | 0.385 |
| RAfu90 | Cochliobolus | 0.406 | 0.846 |
| RAfu91 | Acremonium | 0.334 | 0.61 |
| RAfu92 | Meyerozyma | 0.365 | 0.641 |
| RAfu93 | Phialophora | 0.687 | 0.282 |
| RAfu94 | Trichothecium | 0.169 | 0.196 |
| RAfu95 | Lopharia | 0.878 | 0.926 |
| RAfu96 | Eutypella | 0.753 | 0.719 |
| RAfu97 | Hypochnicium | 0.188 | 0.817 |
| RAfu98 | Xylaria | 0.42 | 0.278 |
| RAfu99 | Physalacria | 0.581 | 0.963 |
| RAfu100 | Stagonospora | 0.945 | 0.186 |
| RAfu101 | Yamadazyma | 0.268 | 0.078 |
| RAfu102 | Microsphaeropsis | 0.512 | 0.463 |
| RAfu103 | Cercospora | 0.54 | 0.319 |
| RAfu104 | Trichaptum | 0.577 | 0.593 |
| RAfu105 | Meira | 0.79 | 0.382 |
| RAfu106 | Oudemansiella | 0.039 | 0.149 |
| RAfu107 | Flavodon | 0.018 | 0.012 |
| RAfu108 | Sistotremastrum | 0.55 | 0.205 |
| RAfu109 | Sarcinomyces | 0.541 | 0.862 |
| RAfu110 | Gymnopilus | 0.36 | 0.086 |
| RAfu111 | Debaryomyces | 0.087 | 0.675 |
| RAfu112 | Coriolopsis | 0.364 | 0.479 |
| RAfu113 | Microdochium | 0.186 | 0.122 |
| RAfu114 | Plectosphaerella | 0.618 | 0.735 |
| RAfu115 | Psilocybe | 0.032 | 0.027 |
| RAfu116 | f__Botryobasidiaceae_Ur | 0.392 | 0.526 |
| RAfu117 | Lophiostoma | 0.252 | 0.066 |
| RAfu118 | f__Trichocomaceae_Uncla | 0.68 | 0.754 |
| RAfu119 | Phlebia | 0.374 | 0.988 |
| RAfu120 | Corticium | 0.029 | 0.282 |
| RAfu121 | Sporobolomyces | 0.908 | 0.374 |
| RAfu122 | Basidiobolus | 0.075 | 0.644 |
| RAfu123 | Pluteus | 0.27 | 0.269 |
| RAfu124 | Parasympodiella | 0.718 | 0.744 |
| RAfu125 | Paraphaeosphaeria | 0.935 | 0.867 |
| RAfu126 | Hymenochaete | 0.3 | 0.733 |
| RAfu127 | Camillea | 0.52 | 0.239 |
| RAfu128 | Exidiopsis | 0.322 | 0.07 |
| RAfu129 | Gerronema | 0.585 | 0.755 |
| RAfu130 | Waitea | 0.23 | 0.136 |
| RAfu131 | Daldinia | 0.617 | 0.997 |
| RAfu132 | Ramichloridium | 0.773 | 0.34 |
| RAfu133 | Tetraplosphaeria | 0.664 | 0.344 |
| RAfu134 | Microporus | 0.422 | 0.685 |
| RAfu135 | Nigroporus | 0.426 | 0.542 |
| RAfu136 | f__Elsinoaceae_Unclassi | 0.908 | 0.795 |
| RAfu137 | Monographella | 0.041 | 0.233 |
| RAfu138 | Eichleriella | 0.878 | 0.681 |
| RAfu139 | Trichoderma | 0.055 | 0.353 |
| RAfu140 | Hypoxylon | 0.987 | 0.434 |
| RAfu141 | Paecilomyces | 0.831 | 0.295 |
| RAfu142 | Valsa | 0.021 | 0.109 |
| RAfu143 | Botryosphaeria | 0.15 | 0.201 |
| RAfu144 | Vascellum | 0.532 | 0.32 |
| RAfu145 | Leptoxyphium | 0.083 | 0.237 |
| RAfu146 | Phaeococcus | 0.418 | 0.451 |
| RAfu147 | f__Polyporaceae_Unclass | 0.971 | 0.666 |
| RAfu148 | Polyporus | 0.116 | 0.315 |
| RAfu149 | Zasmidium | 0.64 | 0.444 |
| RAfu150 | Poitrasia | 0.311 | 0.542 |
| RAfu151 | Parasola | 0.763 | 0.635 |
| RAfu152 | Inonotus | 0.796 | 0.538 |
| RAfu153 | Pyrenochaeta | 0.225 | 0.545 |
| RAfu154 | Chaetomium | 0.569 | 0.385 |
| RAfu155 | Dissoconium | 0.765 | 0.668 |
| RAfu156 | Bullera | 0.669 | 0.52 |
| RAfu157 | f__Microbotryaceae_Uncl | 0.298 | 0.234 |
| RAfu158 | Polychaeton | 0.386 | 0.357 |
| RAfu159 | Dirinaria | 0.546 | 0.372 |
| RAfu160 | Passalora | 0.249 | 0.396 |
| RAfu161 | Preussia | 0.685 | 0.271 |
| RAfu162 | Xenasmatella | 0.292 | 0.2 |
| RAfu163 | Thielaviopsis | 0.331 | 0.234 |
| RAfu164 | Magnaporthe | 0.754 | 0.625 |
| RAfu165 | Annulohypoxylon | 0.514 | 0.601 |
| RAfu166 | Moesziomyces | 0.228 | 0.363 |
| RAfu167 | f__Arthopyreniaceae_Unc | 0.046 | 0.018 |
| RAfu168 | Tilletiopsis | 0.943 | 0.597 |
| RAfu169 | Orbilia | 0.202 | 0.48 |
| RAfu170 | Dendryphiella | 0.837 | 0.903 |
| RAfu171 | Phialemonium | 0.863 | 0.728 |
| RAfu172 | Humicola | 0.657 | 0.865 |
| RAfu173 | Coniothyrium | 0.832 | 0.157 |
| RAfu174 | Meripilus | 0.943 | 0.912 |
| RAfu175 | Stilbohypoxylon | 0.958 | 0.629 |
| RAfu176 | Chaetomella | 0.141 | 0.592 |
| RAfu177 | Fomitopsis | 0.534 | 0.693 |
| RAfu178 | Fusidium | 0.08 | 0.021 |
| RAfu179 | f__Leptosphaeriaceae_Ur | 0.016 | 0.009 |
| RAfu180 | Hannaella | 0.4 | 0.36 |
| RAfu181 | Kazachstania | 0.751 | 0.41 |
| RAfu182 | Sporisorium | 0.088 | 0.52 |
| RAfu183 | Myrmecridium | 0.956 | 0.39 |
| RAfu184 | Sterigmatomyces | 0.454 | 0.168 |
| RAfu185 | Brycekendrickomyces | 0.471 | 0.964 |
| RAfu186 | Perenniporia | 0.248 | 0.763 |
| RAfu187 | f__Massarinaceae_Unclas | 0.694 | 0.76 |
| RAfu188 | Stenella | 0.542 | 0.614 |
| RAfu189 | f__Phaeosphaeriaceae_Ur | 0.55 | 0.386 |
| RAfu190 | Torulaspora | 0.007 | 0.005 |
| RAfu191 | f__Bionectriaceae_Uncla | 0.784 | 0.934 |
| RAfu192 | Verrucaria | 0.639 | 0.444 |
| RAfu193 | Sydowia | 0.104 | 0.911 |
| RAfu194 | Sporidiobolus | 0.194 | 0.812 |
| RAfu195 | Pichia | 0.563 | 0.293 |
| RAfu196 | Camarosporium | 0.343 | 0.277 |
| RAfu197 | Rhizopus | 0.925 | 0.797 |
| RAfu198 | Gymnopus | 0.192 | 0.207 |
| RAfu199 | Hypholoma | 0.119 | 0.171 |
| RAfu200 | Capnodium | 0.145 | 0.691 |
| RAfu201 | Glomerella | 0.702 | 0.926 |
| RAfu202 | Echinochaete | 0.181 | 0.357 |

TableS9 abundance identified ta

| Genus | Presence frequency (n=21) | Range of relative abundance (%) |
|---|---|---|
| Bacteria | | |
| *Sphingobium* | 21 | 0.02-0.15 |
| *Rhodomicrobium* | 17 | 0-0.13 |
| *Shimwellia* | 21 | 0.01-1.65 |
| *Izhakiella* | 5 | 0-0.13 |
| *Solirubrobacter* | 19 | 0-0.04 |
| *Pleurocapsa _PCC-7327* | 21 | 0.01-0.26 |
| *Robinsoniella* | 13 | 0-0.11 |
| *JGI_0001001_H03* | 11 | 0-0.02 |
| Fungi | | |
| *Torulaspora* | 7 | 0-0.01 |
| f_*Leptosphaeriaceae* g_unidentified | 7 | 0-0.12 |